\documentclass[fleqn,usenatbib]{mnras}
\usepackage[T1]{fontenc}
\usepackage{ae,aecompl}

\usepackage{graphicx}
\usepackage{amsmath}
\usepackage{amssymb}

\def\ba#1\ea{\begin{align}#1\end{align}}
\newcommand{\bit}{\begin{itemize}}
\newcommand{\eit}{\end{itemize}}

\newcommand{\nn}{\nonumber}

\title[Parity wise alignments of CMB multipoles]{Alignments of parity even/odd-only multipoles in CMB}

\author[P. K. Aluri et al.]{
Pavan K. Aluri,$^{1}$\thanks{E-mail: aluripavan@gmail.com}
John P. Ralston,$^{2}$\thanks{Email: ralston@ku.edu}
Amanda Weltman$^{1,3,4}$\thanks{Email: amanda.weltman@uct.ac.za}
\\
$^{1}$Cosmology \& Gravity Group, Department of Mathematics and Applied Mathematics, University of Cape Town, Rondebosch 7700, South Africa\\
$^{2}$Department of Physics and Astronomy, University of Kansas, Lawrence, KS 66045, USA\\
$^{3}$Institute for Advanced Study, Princeton, NJ 08540, USA\\
$^{4}$Center for Computational Astrophysics, Flatiron Institute, 162 Fifth Avenue, New York, NY, USA
}

\date{Accepted XXX. Received YYY; in original form ZZZ}

\pubyear{2017}

\begin{document}
\label{firstpage}
\pagerange{\pageref{firstpage}--\pageref{lastpage}}
\maketitle

\begin{abstract}
We compare the statistics of parity even and odd multipoles of the cosmic microwave background (CMB) sky from PLANCK full mission temperature measurements. An excess power in odd multipoles compared to even multipoles has previously been found on large angular scales. Motivated by this apparent parity asymmetry, we evaluate directional statistics associated with even compared to odd multipoles, along with their significances. Primary tools are the \emph{Power Tensor} and \emph{Alignment Tensor} statistics. We limit our analysis to the first sixty multipoles i.e., $l=[2,61]$. We find no evidence for statistically unusual alignments of even parity multipoles. More than one independent statistic finds evidence for alignments of anisotropy axes of odd multipoles, with a significance equivalent to $\sim 2 \sigma$ or more. The robustness of alignment axes is tested by making galactic cuts and varying the multipole range. Very interestingly, the region spanned by the (a)symmetry axes is found to broadly contain other parity (a)symmetry axes previously observed in the literature.
\end{abstract}




\section{Introduction}
Many tests of symmetry of the cosmic microwave background (CMB) sky have revealed unexplained anomalies on large angular scales, namely among low multipoles. Many low multipoles are plagued with anomalous features, associated with a breakdown of isotropy, with significances that varied between different data releases \citep{Tegmark04, RalstonJain04, Copi04, Schwarz04, Eriksen04, Akrami14, Vielva04, Land05a, KimPavel10, MirrorParity}.  In some cases the anomalies have been attributed to statistical flukes.  However, they have received significant interest from the cosmology community by way of alternate or independent analyses towards understanding these peculiarities (see for example
\cite{Hajian03,Slosar05,Land05b,Bielewicz05,Costa06,Copi06,Wiaux06,Abramo06,Bernui07,Gruppuso09,Sarkar11,
Cruz11,Rassat13,Rassat14,Polastri15,Copi15,Schwarz16,Gruppuso11,Hansen11,Kim11,Maris11,Aluri12,Naselsky12,
Eriksen07,Bernui08,Lew08,Hansen09,Bernui09,Paci10,Santos12,Flender13,Rath13,Bernui14,Quartin15,Aiola15,
Gurzadyan09,Naselsky11,BenDavid12,Cruz06,Cruz07,Nadathur14}). Regardless of interpretation, the large scale anomalies have persisted from WMAP to PLANCK mission data, where the science teams pursued them with no final conclusion \citep{WMAP7yrAnom,wmap9yrfinalres,plk13is,plk15is}.

In this paper we uncover yet another peculiarity associated with low multipole CMB data. We compare the alignments of parity-even and parity-odd multipoles separately to explore any preferred directions associated with each. The significances of these directions point to a particular parity preference present in the data, and possible clues about their relation to other large angle CMB anomalies.

In~\citet{KimPavel10}, an anomalous  point (inversion) parity asymmetry was reported to be present in CMB data at low$-l$. Odd multipoles of the CMB were found to have significantly more power compared to the even multipoles in the angular power spectrum from WMAP seven year data, following an earlier analysis that used WMAP first year data \citep{Land05a}.
Let $P^+ = \langle D_l \rangle_{even-l}$ and $P^- = \langle D_l \rangle_{odd-l}$ denote mean power in even and odd multipoles, respectively, up to a chosen $l_{max}$ in the multipole range $l=[2,l_{max}]$. Here $D_l = l(l+1)C_l/2\pi$, and $C_l$ is the CMB angular power spectrum.
Since the power $l(l+1)C_l \sim$~constant, at low multipoles, the ratio $R(l_{max})=P^+/P^-$ is expected to fluctuate about `$1$'. However it was found to be significantly lower than `$1$' with a probability-to-exceed the observed value in data reaching a minimum of $\sim 3\sigma$ at $l_{max}=22$.

This was followed by other studies confirming the anomalous nature of
this parity asymmetry \citep{Gruppuso11, Aluri12}. In the PLANCK 2015
analysis \citep{plk15is}, the $p-$value of this asymmetry was evaluated
to be $0.2-0.3\%$ at $l_{max}=28$, depending on the specific component
separation method used to extract the CMB signal.

The directionality of this parity asymmetry was probed by~\citet{WZhao}, where
the ratio $R(l_{max})$ and its variants were computed in different sky directions
to obtain a map of the even-odd power asymmetry with a chosen $l_{max}$. Curiously,
the minimum of the odd parity excess statistic, $R(l_{max})$, was found to
occur in the direction of the CMB dipole.

Here we analyse the even and odd multipoles separately in a wider multipole
range, to explore any preferred directions associated with these point parity
(a)symmetry modes.

\section{Power tensor, Power entropy, and Alignment Statistics}
The Power tensor is a robust diagnostic to test isotropy 
of CMB data \citep{RalstonJain04,Samal08,Samal09}.  
The CMB temperature is conventionally expanded in terms of spherical
harmonics $Y_{lm}(\hat{n})$: 
\begin{equation}
\Delta T (\hat{n}) = \sum_{l=2}^{\infty} \sum_{m=-l}^{+l} a_{lm} Y_{lm}(\hat{n})\,.
\end{equation}
Here $a_{lm}$ are the spherical harmonic coefficients, $\Delta T(\hat{n})$
denotes the CMB temperature anisotropies after subtracting the monopole and
dipole, and $\hat{n}$ is the position vector on the dome of the sky.

In Dirac notation, the coefficients of the spherical harmonic expansion are
\begin{equation}
a_{lm} = \langle l,m| a(l)\rangle\,,
\end{equation}
where $|l,m\rangle$ represent eigenstates of the angular momentum operators
$J^2$ and $J_z$.
Under a small rotation, the $a_{lm}$'s change to
\begin{equation}
|a(l)\rangle' = |a(l)\rangle + |\delta a(l) \rangle\,,
\end{equation}
where the infinitesimal change is given by $|\delta a(l) \rangle\ = -i {\bf J} \cdot {\bf \Theta} |a(l)\rangle$.
Here $J_i$ ($i=1\cdots3$) are the angular momentum matrices in spin$-l$
representation, and $\Theta_i$ are the angles of rotation.
To find the axes along which the maximum change is achieved, compute
the Hessian, which is
\begin{equation}
\frac{\partial}{\partial \Theta_i \partial \Theta_j}  \langle \delta a(l) | \delta a(l)\rangle
 = \langle a(l) | J^i J^j | a(l) \rangle \equiv A_{ij}\,.
\end{equation}
The eigenvectors of $A_{ij}$ define the frame to which maximal change is developed
under rotations.
The corresponding statistic $A_{ij}$ that we call \emph{Power tensor},
is defined as
\begin{equation}
A_{ij} (l) = \frac{1}{l(l+1)(2l+1)} \sum_{mm'm''} a_{lm} J^i_{mm'} J^j_{m'm''} a^*_{lm''}\,.
\label{eq:pt}
\end{equation}
Under the assumption of statistical isotropy, different spherical harmonic coefficients are
uncorrelated i.e., $\langle a_{lm} a^*_{l'm'} \rangle = C_l \delta_{ll'} \delta_{mm'}$,
and hence $\langle A_{ij} \rangle = (C_l/3) \delta_{ij}$.
Thus, in an ensemble realizations of an uncorrelated, statistically isotropic CMB sky,
the eigenvalues of the Power tensor are randomly distributed about the mean value of $C_l/3$.
The Power tensor  eigenvectors are also distributed uniformly over the sky.

Thus, Power tensor
maps the complicated pattern of each multipole on the sky onto an ellipsoid whose
axes lengths are given by its eigenvalues, and the three ellipsoid
axes by its eigenvectors.
Hence, Power tensor can be used to characterise axiality, planarity, as well as consistency
with isotropy of each multipole by comparing the ratio of its eigenvalues (ie., shape
of the ellipsoid), with the
corresponding eigenvector denoting the direction of isotropy breakdown.

In any given realization, the eigenvalues of the Power tensor will not
be  equal.  Let the eigenvalues and eigenvectors corresponding to a multipole be denoted
$\Lambda_\alpha$ and $e^i_\alpha$, where `$\alpha$' denotes the three eigen-indices
and `$i$' denotes the components of each eigenvector ${\bf e}_\alpha$. We also define the
{\it principal eigenvector} (PEV) as the eigenvector associated with the largest eigenvalue.
Each PEV is then taken as the anisotropy axis corresponding to a multipole $l$.

The significance of anisotropy/axiality represented by a PEV can be quantified using 
\emph{Power entropy}, defined as
\begin{equation}
S = - \sum_{\alpha=1}^{3} \lambda_\alpha \ln(\lambda_\alpha)\,,
\end{equation}
where $\lambda_\alpha = \Lambda_\alpha/\sum_\beta \Lambda_\beta$ are the normalized
eigenvalues of the Power tensor.
In the limit that a multipole is highly anisotropic, one normalized eigenvalue will
tend to being `$1$'. Correspondingly, the Power entropy
$S\rightarrow 0$. If statistical isotropy holds, then each normalized eigenvalue is
equal to $1/3$, and $S \rightarrow \ln(3) \approx 1.0986$, which is the maximum possible value.

The PEV's make it possible to compare the {\it orientations} of different multipoles,
which {\it a priori} contain information, that is independent of the power. A typical statistic
is the dot-product-squared of PEV's from two distinct multipoles $l$ and $l'$.
Squaring the dot-product removes the arbitrary sign convention of eigenvectors.

To quantify correlations in a set of PEV's from a range
$l_{min} \leq l \leq l_{max}$, we use the {\it Alignment tensor} $X$,
which is defined as
\begin{equation}
X_{ij}(l_{min}, \,l_{max}) =\sum_{l=l_{min}}^{l_{max}} \tilde{e}_l^i \tilde{e}_l^j\,,
\label{eq:at} 
\end{equation}
where $\tilde{\bf e}_l$ is the principal eigenvector of a multipole $l$.
Let $\zeta_\alpha$ and ${\bf f}_\alpha$ denote the normalized eigenvalues and
eigenvectors of this Alignment tensor.
The eigenvalues are normalized to remove the trivial effect of the $l$-range.
One then computes the {\it Alignment entropy}, $S_{X}$, which is a rotationally invariant summary of
the ratios of $ \zeta_\alpha$, that  is given by
\begin{equation}
S_X = - \sum_{\alpha=1}^3 \zeta_\alpha \ln(\zeta_\alpha)\,.
\label{eq:as}
\end{equation}

When the PEV's over the range are uncorrelated, $X_{ij} \sim \delta_{ij}$
and all $ \zeta_\alpha$ are equal. In the extreme opposite case when the PEV's over the set of multipoles are all parallel
to a single eigenvector,
then all but one  $ \zeta_\alpha \rightarrow 0$. That leads to the maximal range of Alignment entropy
as $ 0 \leq S_X \leq \ln( 3) $. The lower limit $S_{X} \rightarrow 0$ represents the maximum
possible correlation. The upper limit  $S_{X} \rightarrow \ln(3)$ represents the completely uncorrelated
hypothesis of the standard Big Bang.
We define the {\it collective alignment vector} of a set of multipoles as the principal eigenvector of the corresponding Alignment tensor ($\tilde{\bf f}_\alpha$).  It's significance is assessed using Alignment entropy.
The reader may refer to \cite{RalstonJain04,Samal08} for more details about the Power tensor method,
as well as it's relation to axes inferred from other statistics viz. the angular momentum dispersion
maximization \citep{Tegmark04} and Maxwell's multipole vectors \citep{Schwarz04}.

\section{Description of procedure and data sets}

\subsection{Analysis procedure}

The Power tensor and Alignment tensor allow us to probe any underlying anisotropy axis associated with CMB anisotropies from a desired multipole range or a set of multipoles.

Under point inversion, $a_{lm}\rightarrow (-1)^la_{lm}$, and so
the even(odd) multipoles are symmetric(antisymmetric) under such operation.
In the present work, we apply the Alignment tensor statistic to even and odd multipoles separately. Thus we can explore any common preferred axes underlying these modes separately.

 We first  compute the principal eigenvector (PEV) corresponding to each multipole in a chosen multipole range $l=[l_{min},l_{max}]$. The PEVs are separated  between even and odd multipoles to construct the Alignment tensor for each parity set separately. The PEV of the Alignment tensor will provide the common anisotropy axis corresponding to each set of parity even/odd multipoles under study. The significance of anisotropy represented by this axis is measured using Alignment entropy. This is done by computing the lower tail probability deduced from simulations in comparison to the observed entropy value from data. We also study  alignments in cumulative multipole bins, by varying the upper and lower end of the $l-$range being considered.

\subsection{Real and mock data used}\label{sec:data-sim}
For this study we use the full sky {\tt Commander} CMB map, derived
from PLANCK 2015 data, that is made publicly
available\footnote{\url{http://irsa.ipac.caltech.edu/data/Planck/release_2/all-sky-maps/matrix_cmb.html}}.
It is a maximum likelihood estimate of the CMB map, along with various
astrophysical components such as galactic synchrotron, thermal dust, their
spectral indices, etc., that uses multi-frequency CMB observations
and external observations/templates for various galactic emission types
\citep{EriksenCMDR1,EriksenCMDR2,plk15cmb,plk15fg}.

 The {\tt Commander} map is available at a resolution of {\tt HEALPix}\footnote{\url{http://healpix.jpl.nasa.gov/}} $N_{side}=2048$.
However, we downgrade the map to a lower resolution of $N_{side}=256$,
and smooth it to have a Gaussian beam $FWHM=1^\circ$ (degrees).
Since we are interested in large angular scales, this is sufficient
for our purposes.

 We also prepare the mock data accordingly. The PLANCK collaboration has also provided sets of CMB realizations that have the appropriate instrument effects such as beam smoothing, as well as noise realizations for public use\footnote{\url{http://crd.lbl.gov/cmb-data}}. These are referred to as Full Focal Plane (FFP) simulations. We use the FFP8 and FFP8.1 simulation sets for our purpose. The set FFP8 was an initial release that complements the PLANCK 2015 full mission data release. However due to a slight
mismatch in the theoretical power spectrum of CMB used to generate these realizations, with the angular power spectrum consistent with final PLANCK 2015 cosmological parameters, the CMB realizations were updated with a new set denoted as FFP8.1 that match PLANCK 2015 cosmology
\citep{plk15sim}.
Hence we use simulated CMB skies from the set FFP8.1, but will use the FFP8 realizations for noise.

 The FFP simulations of CMB and noise that are provided, correspond to a specific frequency channel, and are not readily usable. These simulation sets do not constitute individual component separated maps corresponding to various cleaning algorithms used by PLANCK such as {\tt Commander}, {\tt SMICA}, etc., to obtain clean CMB maps from the raw satellite data \citep{plk15cmb}. Thus, to obtain a set of realistic CMB maps, we process this ensemble of multi-channel maps as follows.

 We downgrade all the CMB and noise realizations to a common resolution of
{\tt HEALPix} $N_{side}=256$, and smooth to have a uniform beam resolution of
$FWHM=1^\circ$ (degrees) Gaussian beam. We apply the {\tt HEALPix} facilities {\tt anafast}, {\tt alteralm} and {\tt synfast} in that order to bring them to the afore mentioned common \texttt{HEALPix} resolution and beam smoothing. We used the circularized  beam transfer functions corresponding to each PLANCK frequency channel, that are provided with the second public release of PLANCK data. We then compute the noise rms corresponding to each channel using these smoothed/downgraded realizations. These noise rms maps are used to combine the smoothed/downgraded individual frequency specific CMB and noise realizations through inverse noise variance weighting.
Thus we are considering only the diagonal part of the full covariance matrix that results from beam smoothing. However since we are interested in studying large angular scales, the coadded CMB and effective noise maps thus obtained would sufficiently represent the observed sky.

A set of 1000 CMB and noise Monte Carlo realizations are provided with appropriate instrument and noise characteristics through PLANCK public release~2. Correspondingly we generate 1000 co-added CMB maps with noise from the FFP realizations following this procedure.

\section{Results}

We are interested in any preferred directional correlations associated with even versus odd
multipoles corresponding to large angular scales of the CMB sky.
We use the multipole range $l=[2,61]$ for this study.
Before proceeding we discuss the anomalous alignment of quadrupole and
octopole modes of the CMB seen in WMAP as well as PLANCK data, that have
received considerable attention (see \cite{wmap9yrfinalres,plk13is} for
the assessment of the WMAP and PLANCK collaborations).

\subsection{Quadrupole-Octopole alignment}

The alignment of the quadrupole ($l=2$) and octopole ($l=3$) anisotropy axes as seen in the
PLANCK full mission {\tt Commander} map deserves comment.
The directions inferred from the principal eigenvector (PEV) corresponding to $l=2$ and $3$
multipoles are listed in Table~\ref{tab:l23algn}. Since eigenvectors of the Power tensor are
headless vectors, we report the direction of these axes from only one of the hemispheres.
We find that these two modes are well aligned with a separation of only $\approx 6.2^\circ$ (degrees).
This corresponds to a random chance occurrence probability of $1-\cos(6.2^\circ)\approx0.0058$,
which is close to a $3\sigma$ significance.
Together with the CMB dipole, the quadrupole and octopole modes point
towards the Virgo cluster \citep{RalstonJain04}.
These axes are shown in subsequent plots as some of the reference anisotropy directions
seen in the CMB sky. Note that the alignment of CMB temperature quadrupole and octopole
modes was found to improve by appying any additional corrections such as residual
galactic bias correction \citep{Aluri11} or kinetic quadrupole correction, frequency independent
\citep{Schwarz04} or frequency dependent \citep{Miguel}. The PLANCK 2015 foreground
cleaned maps have the frequency independent kinetic Doppler boost contribution subtracted \citep{plk15cmb}.
Here we used the PLANCK's \texttt{Commander} 2015 CMB map as provided.

\begin{table}
\centering
\begin{tabular}{c  c}
\hline\\
$l$ & $(\ell,b)$ (degrees) \\
\hline\\
2 & ($239.8^\circ$,$57.2^\circ$)\\
3 & ($244.3^\circ$,$63.0^\circ$)\\
\hline
\end{tabular}
\caption{The directions corresponding to the PEVs of $l=2,3$ modes obtained from Power tensor
         statistic are listed here. These axes are headless and the quoted direction
         is from the upper galactic hemisphere. These broadly point towards the CMB
         kinetic dipole direction $(\ell,b)=(264^\circ,48^\circ)$ as shown in subsequent plots.
         They are aligned at a mere separation of $\approx 6.2^\circ$.}
\label{tab:l23algn}
\end{table}

\subsection{Parity alignments}
Using the PEVs computed for each `$l$' from the multipole range of our
interest, we construct the Alignment tensor defined in Eq.~[\ref{eq:at}]
for even and odd multipoles separately.

\begin{figure}
\centering
\includegraphics[width=0.9\columnwidth]{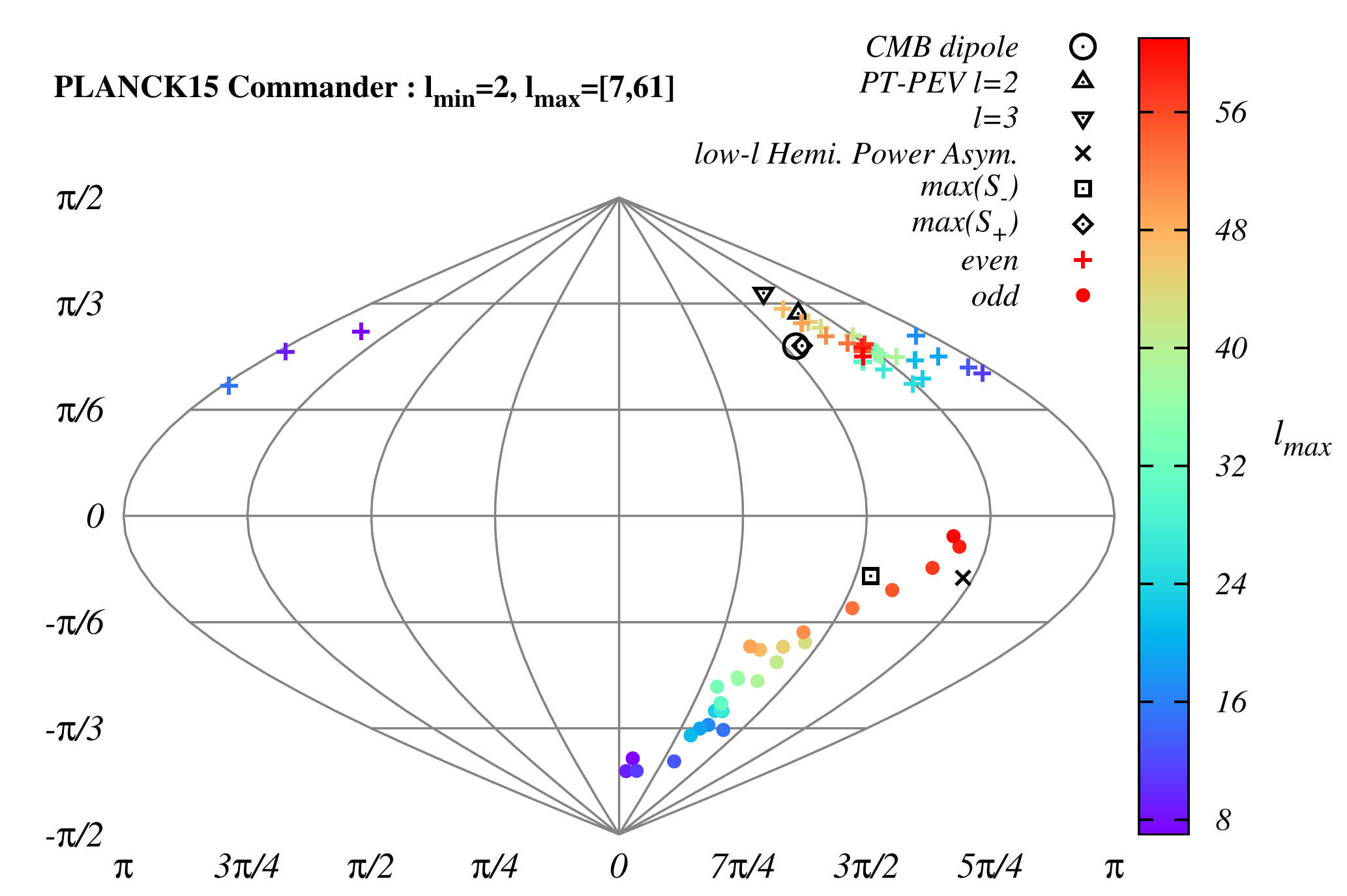}
\caption{Collective alignment vectors i.e., principal eigenvectors of the Alignment tensor
         $X(l_{min}, l_{max})$ (Eq.~\ref{eq:at}) for even and odd multipoles,
         obtained from PLANCK 2015 \texttt{Commander} full sky CMB map are shown
         here in Galactic co-ordinates. The ${\bf +}$'s denote even$-l$ and and the $\bullet$'s
         correspond to odd$-l$ alignment axes. Other prominent anisotropy axes seen in the
         CMB sky are also labeled.}
\label{fig:plk15cmdr-varlmax}
~
~
%
\includegraphics[width=0.9\columnwidth]{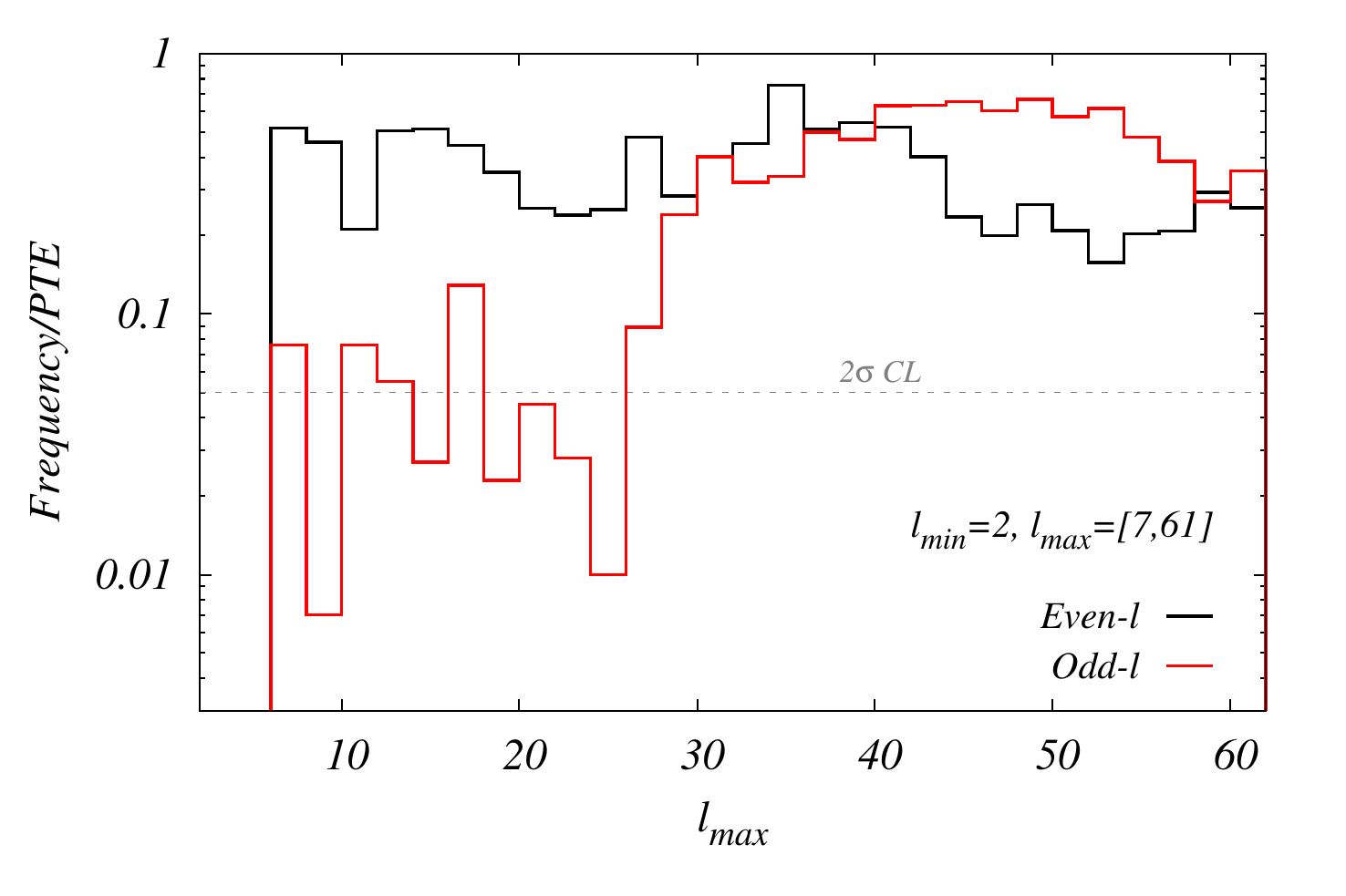}
\caption{Significances of the collective alignment axes, shown in Fig.~[\ref{fig:plk15cmdr-varlmax}],
         as measured using Alignment entropy, $S_X$, are plotted here
         as a function of $l_{max}$. The lower end of the multipole bin is fixed to $l_{min}=2$.
         The upper end of the multipole bin is varied as $l_{max}=[7,61]$.
         The probability to exceed (PTE) the observed value of $S_X$ from data in comparison
         to simulations is plotted in \emph{black} and \emph{red} solid curves for even and odd
         multipoles respectively.
         We see that the odd multipole alignment axes are significantly directional at $\sim 2\sigma$
         level on large angular scales.}
\label{fig:plk15cmdr-varlmax-signf}
\end{figure}

First we present results for the case of varying $l_{max}$, meaning, we fix $l_{min}=2$
and vary $l_{max}=[7,61]$. So, the smallest range considered is $l=2 \cdots 7$, and
the Alignment tensor is computed separately for even and odd multipoles using
$l=2,4,6$ and $l=3,5,7$ PEVs respectively. Then we keep increasing the
multipole range up to $l_{max}=61$ by two multipoles each time (so that there are an 
equal number of even and odd multipoles for computing the Alignment tensor), and
obtain the common anisotropy axis for the set of even/odd multipoles in the current
range every time.
The results are shown in Fig.~[\ref{fig:plk15cmdr-varlmax}].

 There seems to be an apparent clustering of even multipoles, denoted by $+$'s,
broadly oriented along the CMB kinetic dipole ($l=1$) direction. By progressively
adding more multipoles to the Alignment tensor, the derived PEV moves closer to
the CMB dipole direction.
On the other hand, the common alignment axis of odd multipole PEVs, plotted in
the same figure using $\bullet$ point types, steadily drifts from being close
to the southern galactic pole towards the galactic plane.

 We assess the significance of these collective alignment axes of even/odd multipoles using the Alignment entropy ($S_X$) defined in Eq.~[\ref{eq:as}]. The value of the Alignment entropy obtained from the data is compared with the same quantity computed from simulations. The $p-$value plot for the observed value of $S_X$ as a function
of $l_{max}$ is shown in Fig.~[\ref{fig:plk15cmdr-varlmax-signf}]. We find that the
apparent clustering indicated by the common alignment axes of even multipoles (black curve) is not significant, as the $p-$value curve is always within $2\sigma$ in the multipole range considered. However, it could be an indication of a remnant anisotropy (or a leakage)
that is resulting in the apparent clustering of these axes towards CMB dipole.

 In the same plot, Fig.~[\ref{fig:plk15cmdr-varlmax-signf}], we also show the
significances of odd multipole alignment axes as a function of $l_{max}$ (red curve).
We find that these axes are highly directional, despite the
change in their orientation steadily with the addition of more multipoles. The significance fluctuates about the $2\sigma$ confidence level up to $l_{max}=27$, and becomes insignificant thereafter. So, by adding more multipoles, the directionality of
common alignment axis of odd multipoles seen at low$-l$ is weakened.

For reference, we also plot other interesting anisotropy directions seen in the CMB data with different point types in black. The quadrupole and octopole axes listed in Table~\ref{tab:l23algn} of the present analysis are denoted by up and inverted
triangles respectively. The CMB dipole direction, and the low$-l$ hemispherical power asymmetry axis - that is obtained from the analysis of PLANCK 2015 data using the BipoSH framework \citep{plk15is}, are highlighted using a black circle and a cross respectively. A set of interesting anisotropy axes corresponding to a mirror parity (a)symmetry are also found in the CMB data \citep{plk15is}. However, only the mirror asymmetry axis is found to be anomalous. The maximum mirror symmetry axis is labeled $max(S_+)$, and the maximum mirror asymmetry axis is labeled as $max(S_-)$. These two axes are highlighted using a black diamond and a square respectively in Fig.~[\ref{fig:plk15cmdr-varlmax-signf}].

It is interesting to note that the even/odd multipoles' common axes span two broad regions
of the sky in an apparently non-random/non-overlapping manner. One can readily see that the
common alignment axes of even multipole PEVs found here and the (insignificant) even mirror
parity direction - $max(S_+)$, are broadly aligned with the CMB kinetic dipole direction.
The region spanned by the odd multipole common alignment axes contain the odd mirror parity
axis - $max(S_-)$, and the odd parity low$-l$ dipole modulation axis.

\begin{figure*}
\centering
  \includegraphics[width=0.9\columnwidth]{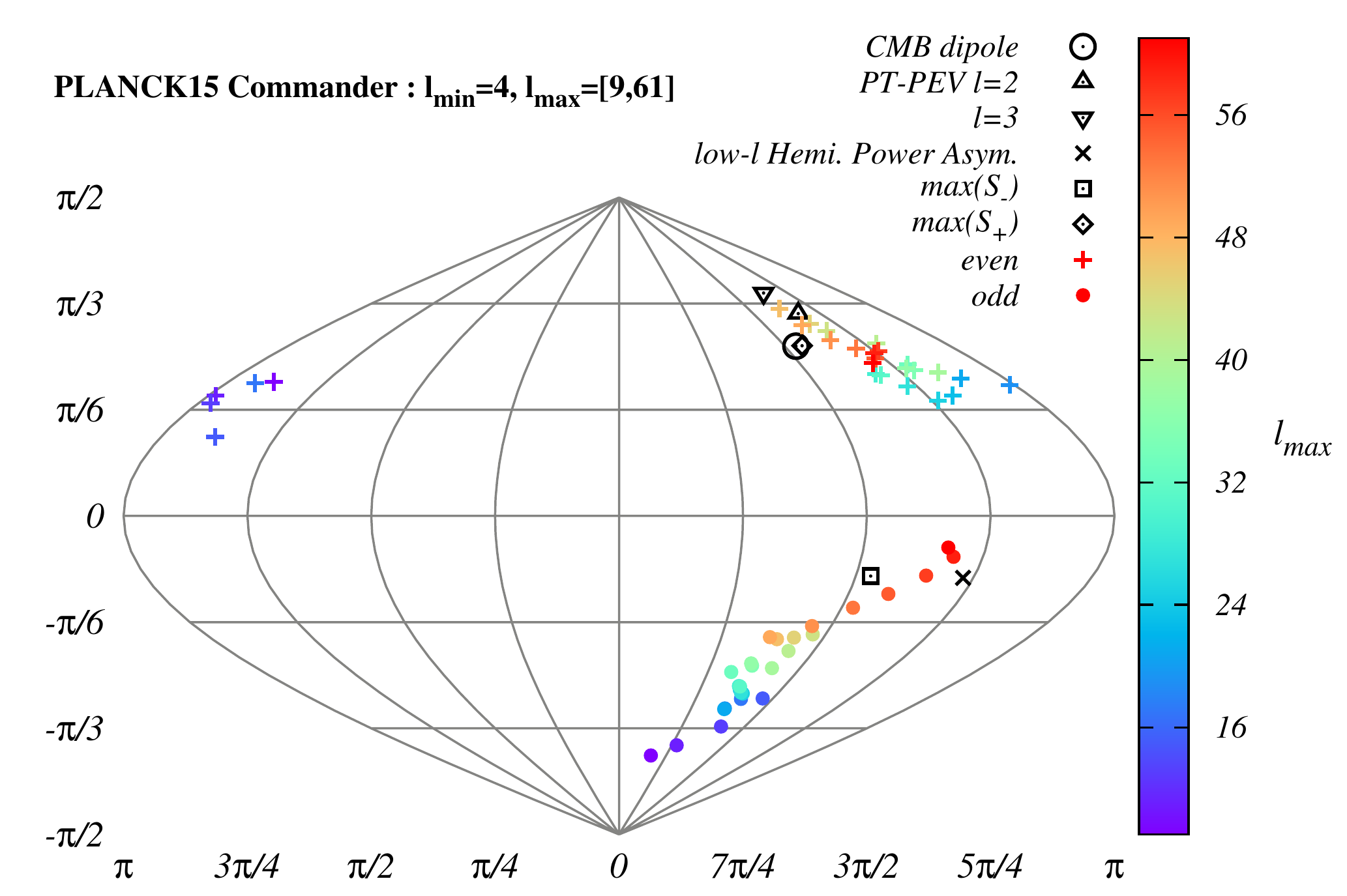}
~
  \includegraphics[width=0.9\columnwidth]{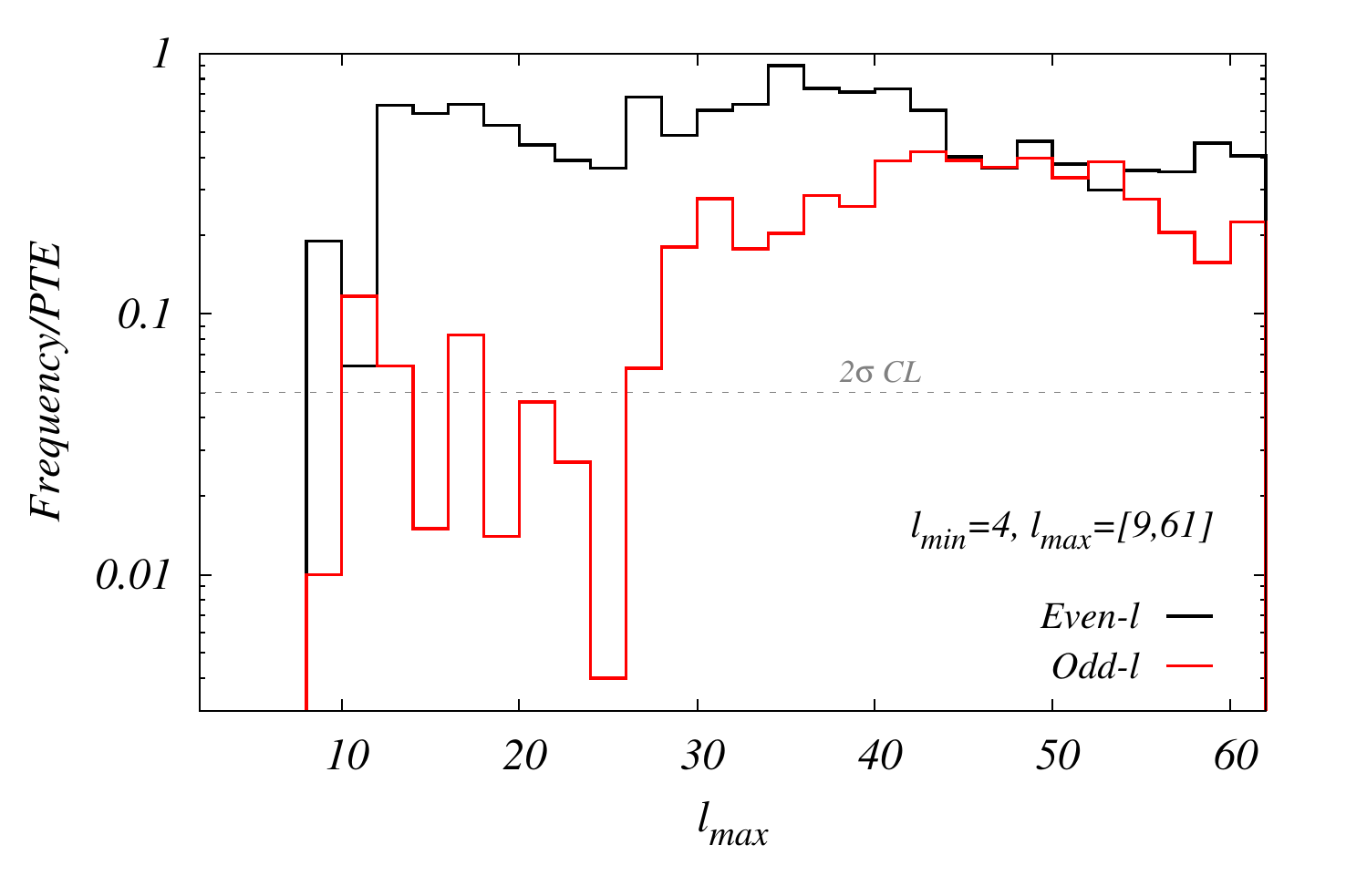}
~
  \includegraphics[width=0.9\columnwidth]{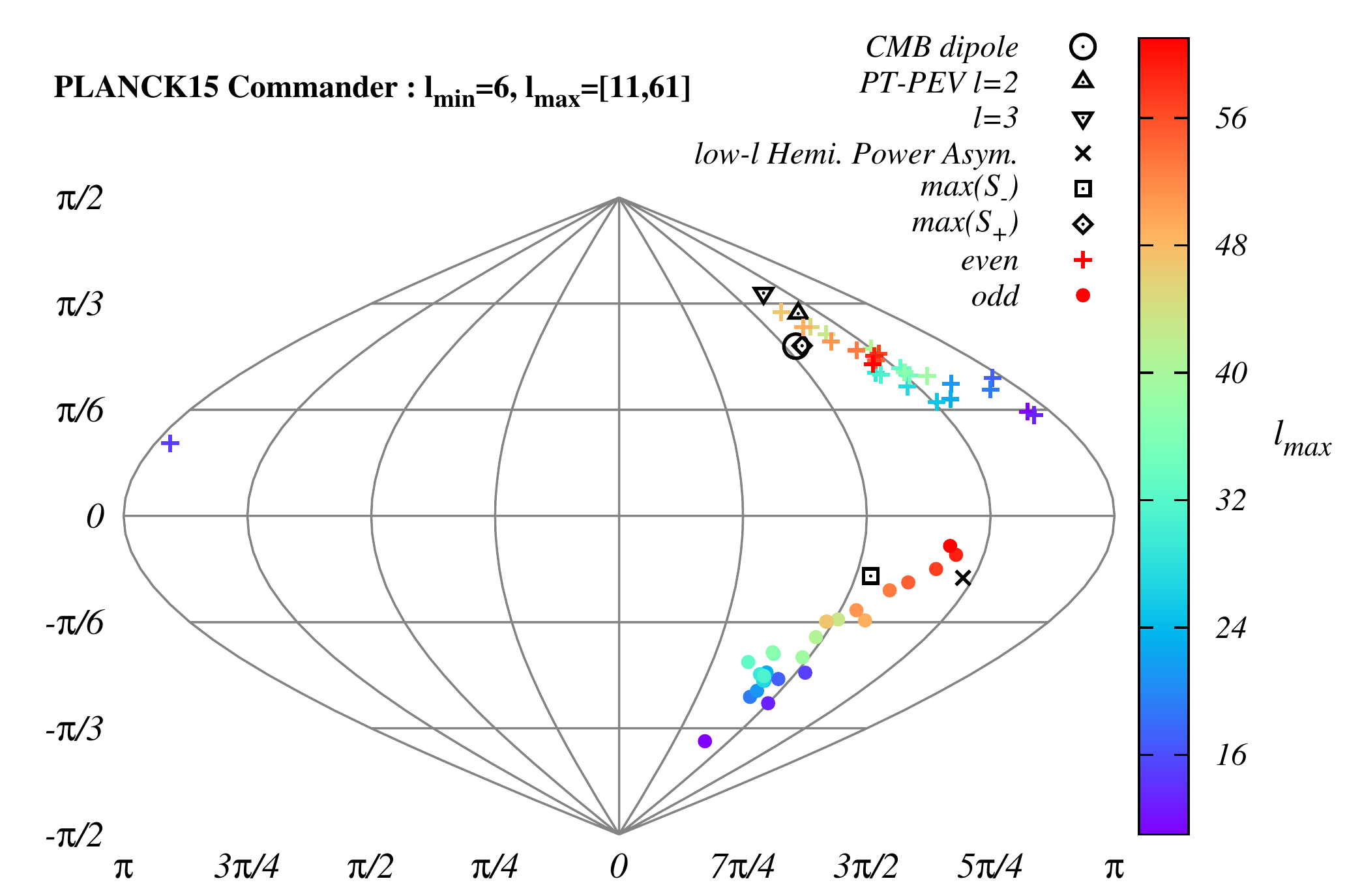}
~
  \includegraphics[width=0.9\columnwidth]{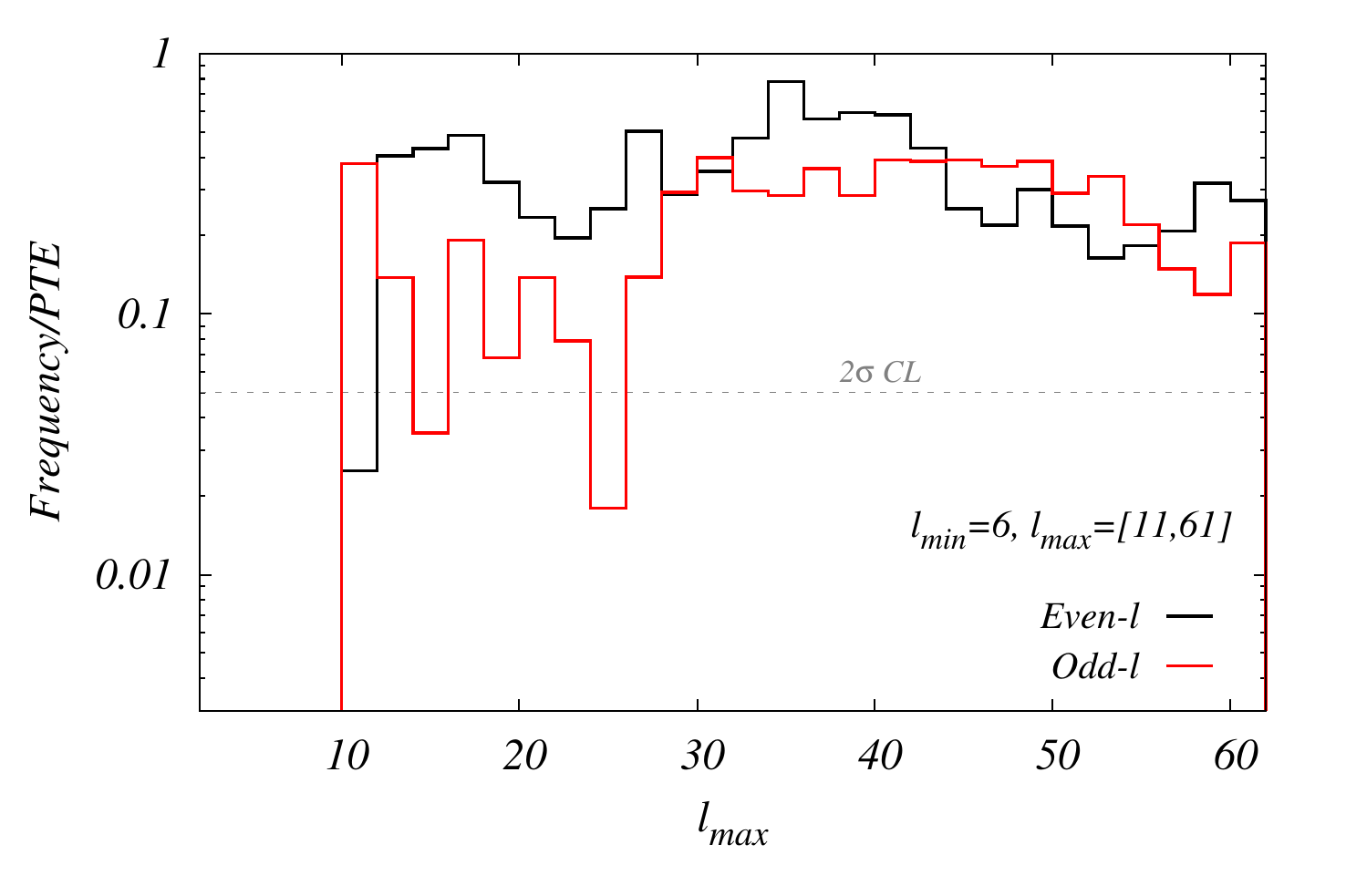}
~
  \includegraphics[width=0.9\columnwidth]{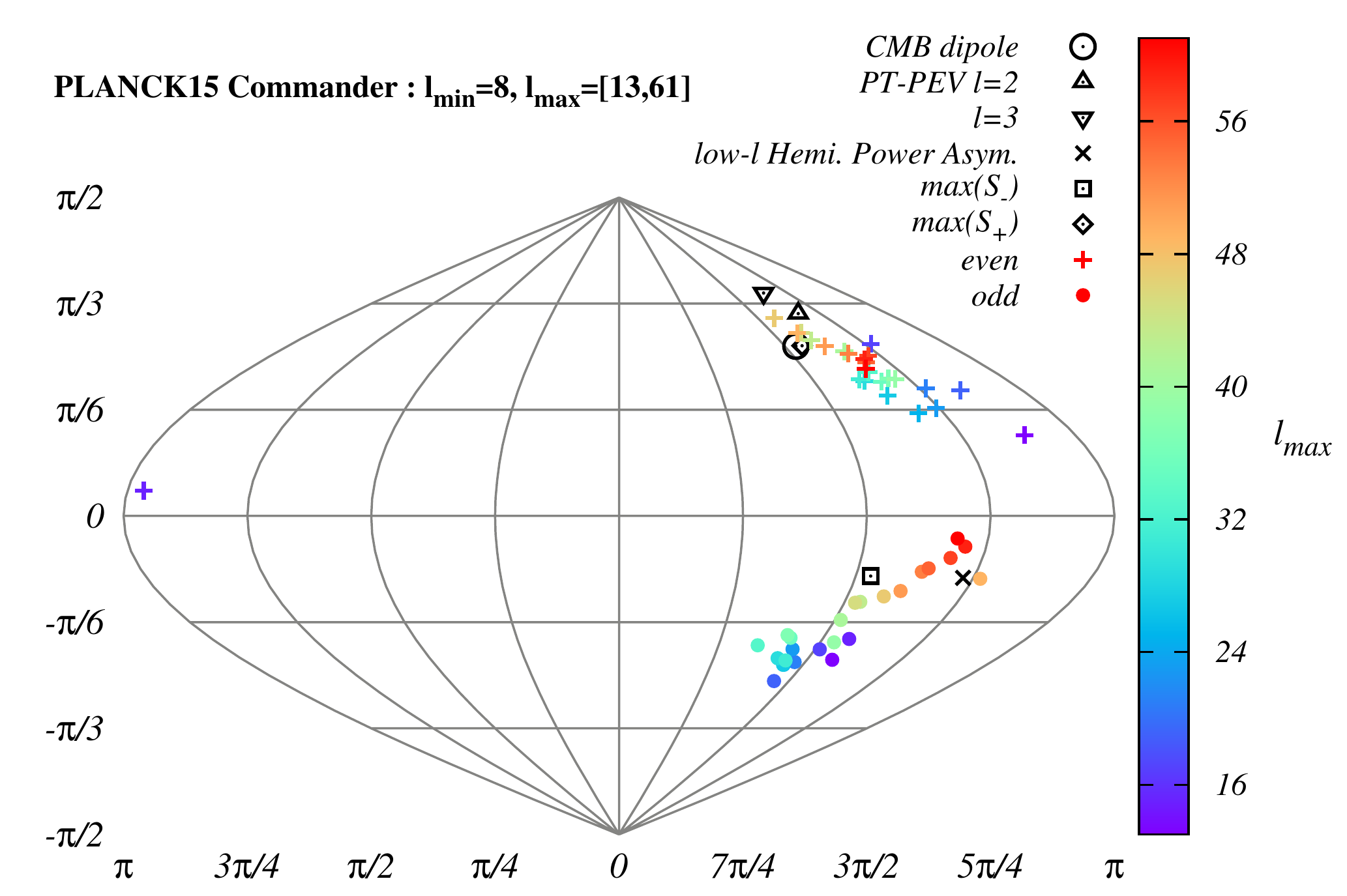}
~
  \includegraphics[width=0.9\columnwidth]{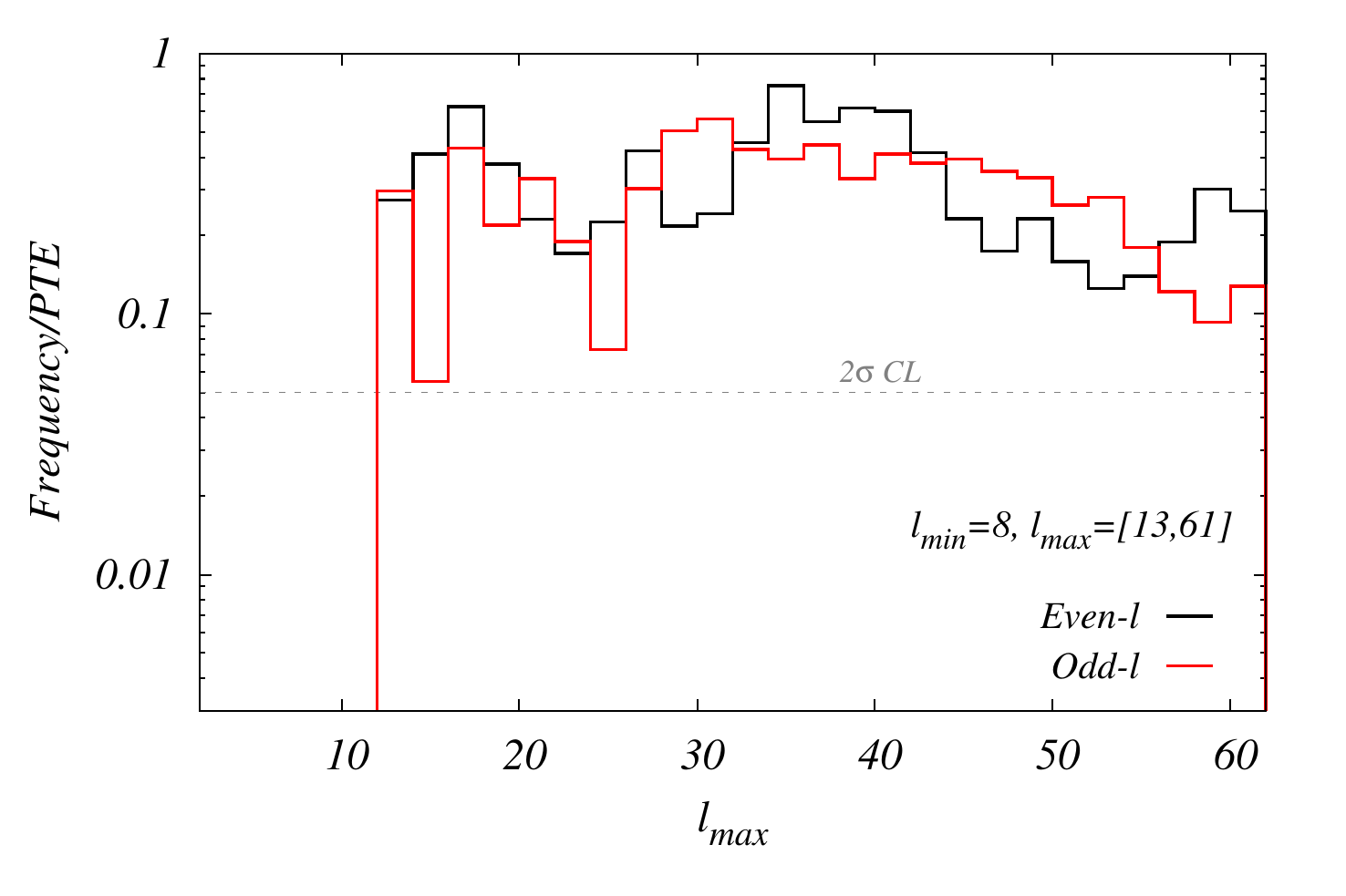}
\caption{Same as Fig.~[\ref{fig:plk15cmdr-varlmax}] and [\ref{fig:plk15cmdr-varlmax-signf}],
         but for different $l_{min}$.
 Although the broad orientation of the axes persists by progressively excluding the first
 few multipoles in these plots, we find that their significances however fall (below $2\sigma$)
as seen from the $p-$value plots shown in \emph{right} column.}
\label{fig:plk15cmdr-varlmax-lcut-signf}
\end{figure*}

\cite{Aluri12} found that the significance of the even-odd multipole power asymmetry
in CMB angular power spectrum significantly decreases when 
the first few multipoles are omitted. We now test for low multipole contributions to the
$\sim 2\sigma$ significance seen for the directionality of odd multipole alignment axes.
We repeat the calculations, while choosing different $l_{min}$ values i.e., $l_{min}=4,6$
and $8$. The results are shown in Fig.~[\ref{fig:plk15cmdr-varlmax-lcut-signf}] in the left column.
We find that the distribution of common alignment axes still persists for different low$-l$
cuts i.e., using different $l_{min}$, but varying the other end of the multipole window
upto $l_{max}=61$.

However, similar to what was observed by \cite{Aluri12}, we find that the
significance of odd multipole PEV alignments quickly disappears when $l_{min}$ of the
multipole window is varied. The $p-$value plots corresponding to choosing different
$l_{min}$ are shown in the right column of Fig.~[\ref{fig:plk15cmdr-varlmax-lcut-signf}].
The even-multipole alignments remain insignificant in this case as well.
 
\begin{figure}
\centering
\includegraphics[width=0.9\columnwidth]{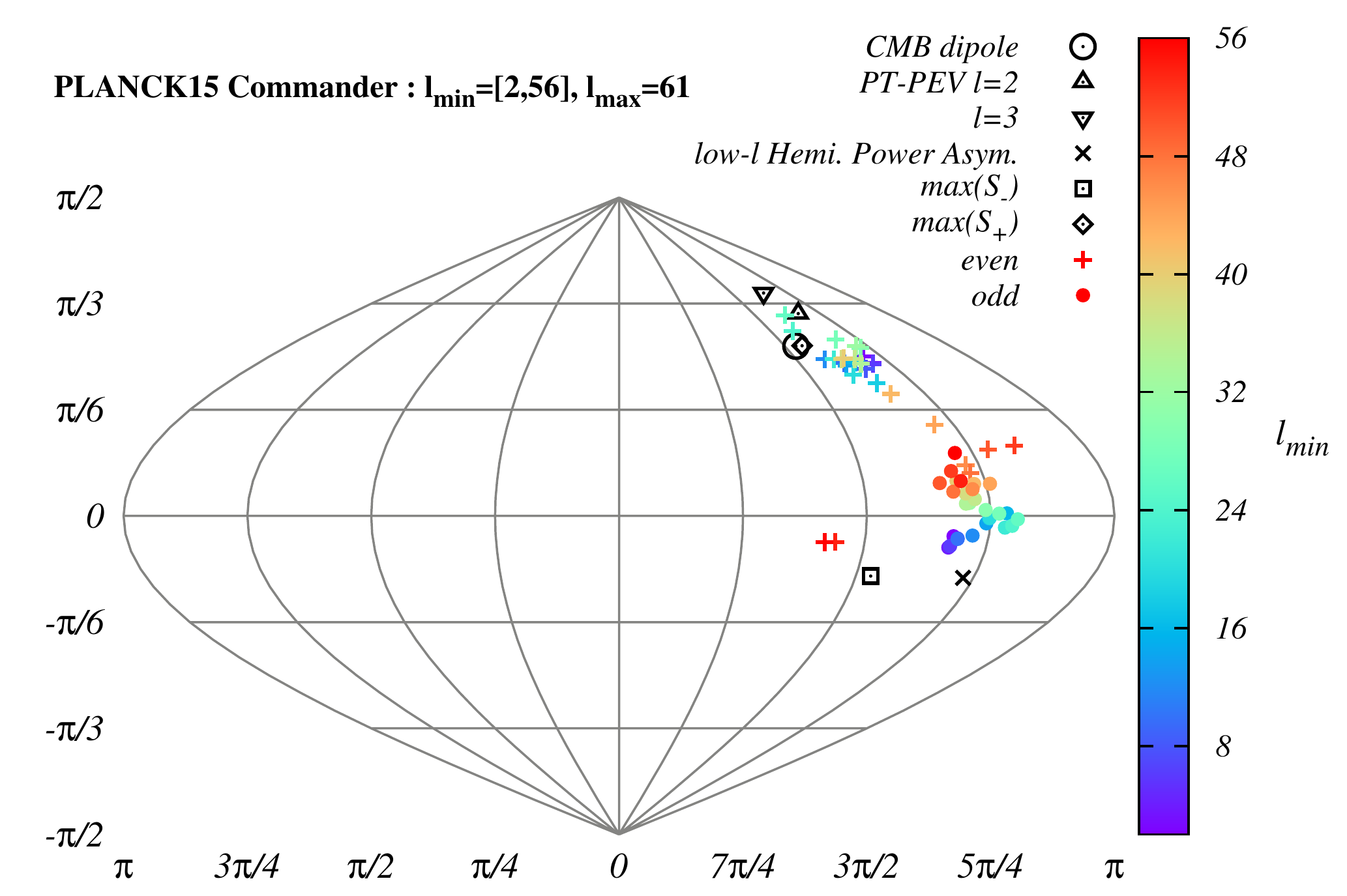}
\caption{The alignment axis of even and odd multipole PEVs (denoted by
         {\bf +} and $\bullet$ respectively), in Galactic co-ordinates, for fixed $l_{max}=61$ and varying
         $l_{min}$ in the range $l=[2,56]$.}
\label{fig:plk15cmdr-varlmin}
\vspace{2em}
%
\includegraphics[width=0.9\columnwidth]{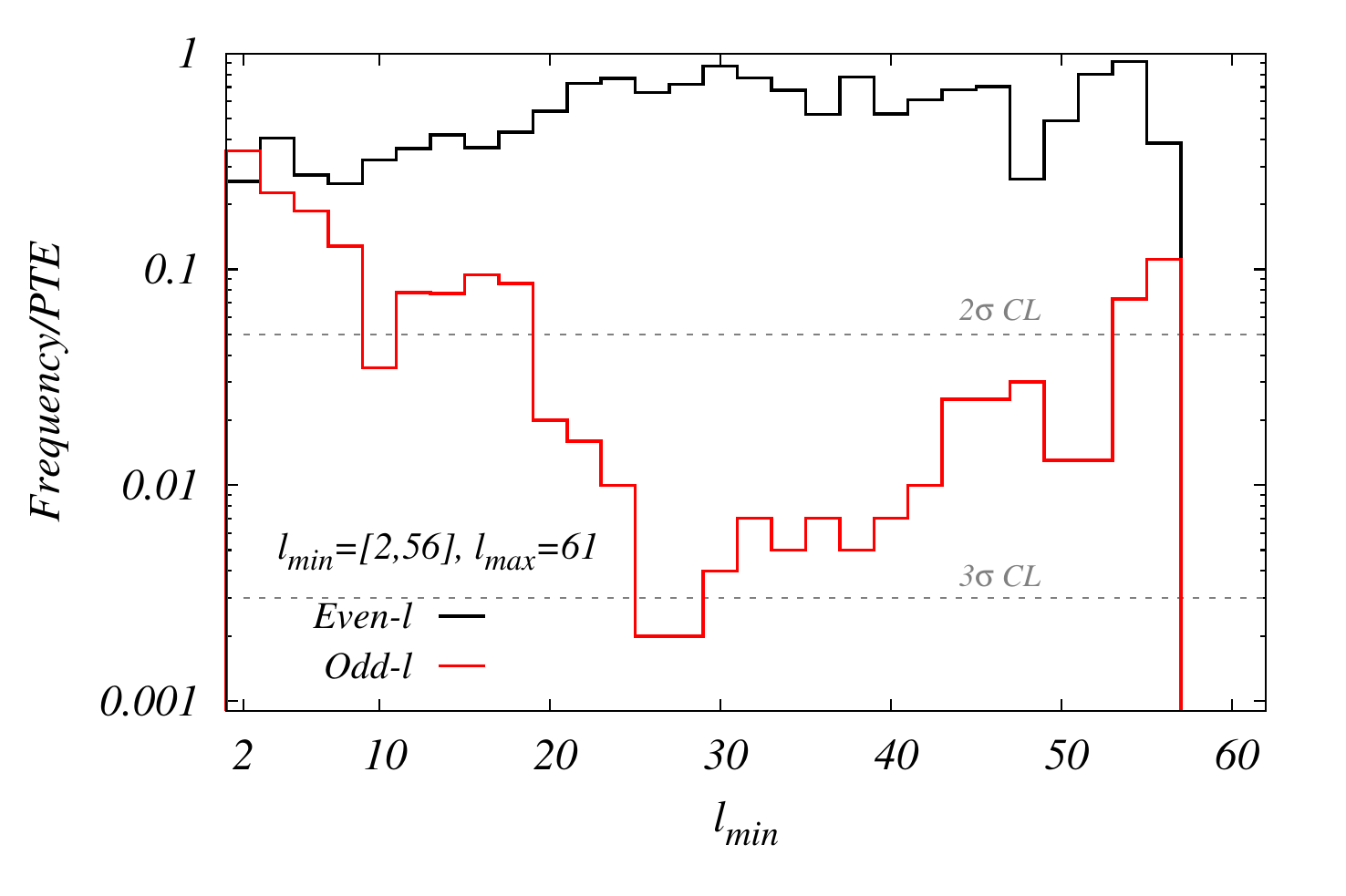}
\caption{The lower tail probabilities or the probability to exceed (PTE) the observed
         Alignment entropy, $S_X$, of the collective alignment axes from data in comparison
         to 1000 simulations as a function of $l_{min}=[2,56]$, while fixing $l_{max}=61$ are shown here.
         The significances of observed $S_X$ of even and odd multipole common anisotropy axes are plotted
         in \emph{black} and \emph{red} solid curves respectively.}
\label{fig:plk15cmdr-varlmin-signf}
\end{figure}

To study the alignment preferences of high$-l$ in the multipole range under consideration, we fix $l_{max}$ and vary $l_{min}$. In Fig.~[\ref{fig:plk15cmdr-varlmin}], we show the collective alignment axes obtained by varying $l_{min}$ in the range $l=[2,56]$, with fixed $l_{max}=61$. The significance of these axes as a function of $l_{min}$ are plotted in Fig.~[\ref{fig:plk15cmdr-varlmin-signf}]. This study suggests a possibility of two distinct populations for $ l \lesssim 30$ compared to $30 \lesssim l \leq 61$ when contrasted with varying $l_{max}$ case. We find a $\sim 2\sigma$ significance upto $l_{max}\sim 30$ in the varying $l_{max}$ case. However in the varying $l_{min}$ case, the significance keeps building up upto $l_{min}\sim 30$ which indicates two distinct populations of anisotropy axes.

 We observe the alignment axis of even multipole PEVs drifting towards the galactic plane as more and more low$-l$ are discarded.
In comparison, the odd multipole PEVs' alignment axis now seem to have settled at the galactic plane. The significance of the common alignment axis becomes acute for $l_{min}\sim 28$. 
A residual foreground bias may explain the clustering of these axes in the galactic
plane, and also the corresponding anomalous significance. We pursue this aspect later in the paper.

\begin{figure*}
\centering
\includegraphics[width=0.42\textwidth]{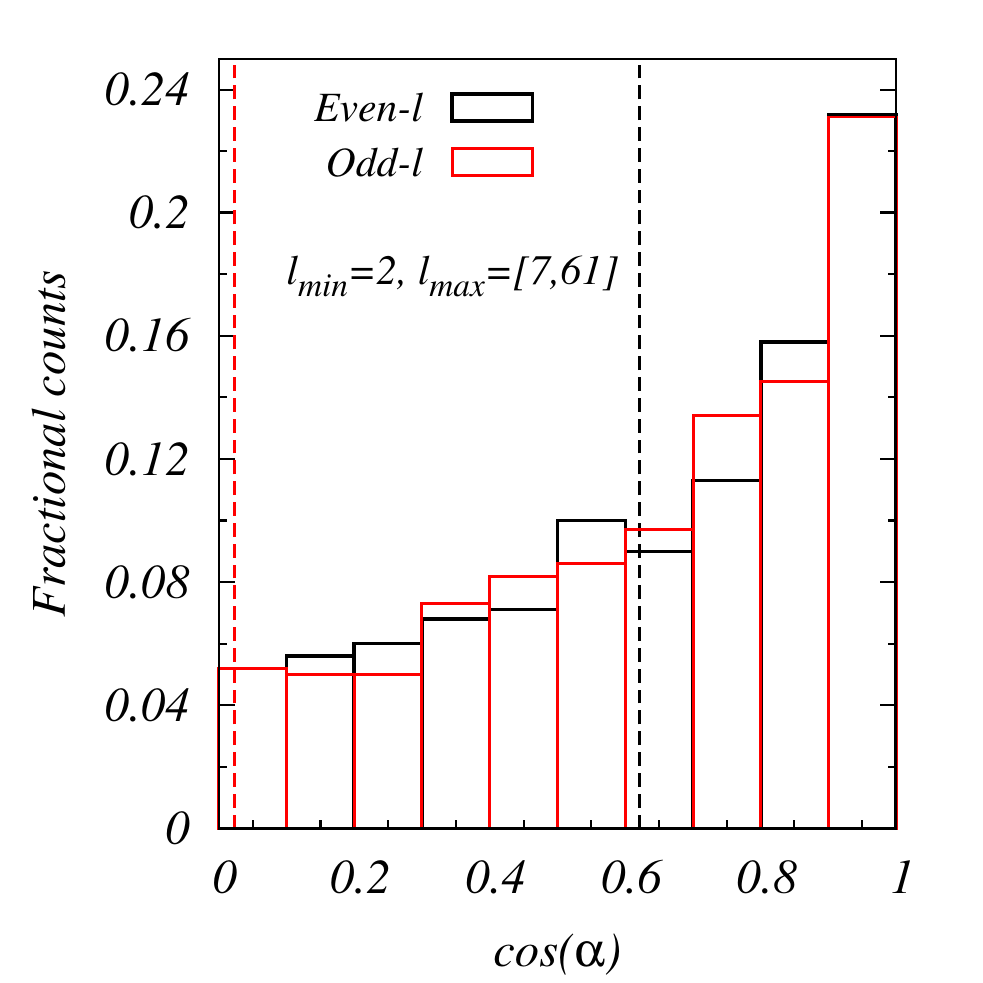}
~
\includegraphics[width=0.42\textwidth]{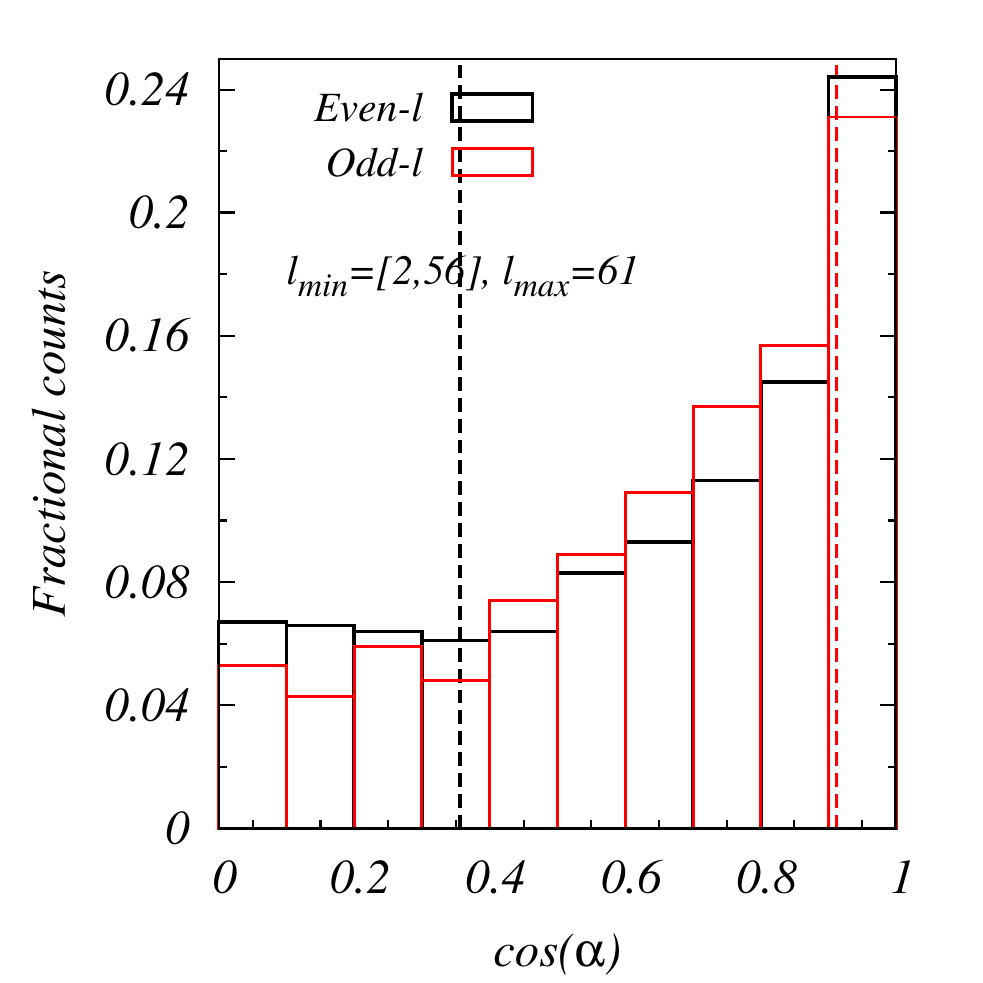}
\caption{Distribution of the observed clustering of even or odd multipole PEV common axes
        computed as dot products of collective alignment axis from the smallest and
         largest multipole bin sets from the range $l=[2,61]$.
         For the varying $l_{max}$ case (\emph{left plot}), the inner product is taken for
         the axes obtained from the multipole bins $l=[2,7]$ and $l=[2,61]$. The varying
         $l_{min}$ case (\emph{right plot}), uses common alignment axes obtained from the
         Alignment tensor for the bins $l=[2,61]$ and $l=[56,61]$.
         The scalar product of collective alignment axes corresponding to smallest and largest bins
         of even/odd multipoles are shown in \emph{black} and \emph{red} solid curves.}
\label{fig:clstr}
\end{figure*}

 Now we probe the observed clustering of common alignment axes of even multipole PEVs  further. The absolute scalar product of the common axes obtained from the smallest and largest subset of multipole bins of even/odd `$l$' PEVs from the whole multipole range $l=[2,61]$ is computed. This product denoted by $\cos(\alpha)$ is taken as representative of these axes being closer or scattered away from each other. The frequency plots of $\cos(\alpha)$ corresponding to even and odd multipoles, as obtained from simulations, are shown in Fig.~[\ref{fig:clstr}].
The two cases of varying $l_{max}$ and $l_{min}$ while fixing the other end of the
multipole window are shown in that figure, in the \emph{left} and \emph{right} panels
respectively. The observed value of the inner product of the same axes from the data are denoted by vertical dashed lines in respective colours. From the histogram plot, we see that the clustering of even multipole common axes is not statistically significant in both cases of varying $l_{max}$ and $l_{min}$. In contrast with this, the scalar product of odd multipoles' common axes from the smallest and largest subsets is statistically significant in comparison to simulations.

 The simulations suggest that the collective alignment axes, computed using Alignment tensor, from the smallest and largest multipole bin windows, tend to be
closer to each other. This could be because the small multipole bin window
is a subset of the larger multipole window, and thus correlated, leading to this preference. Upon extending the multipole window range ($l_{max}$), we observe that the distributions tend towards being uniform, as expected.

We tested the stability of alignment axes by applying galactic masks with different
sky fractions, and inpainting the masked CMB maps using \texttt{iSAP} software\footnote{\url{http://www.cosmostat.org/software/isap/}}
\citep{inpaint}.
The publicly available PLANCK HFI masks were used which exclude  $1\%, 3\%, 10\%$,
$20\%$ and $30\%$ of the sky fraction\footnote{\url{http://irsa.ipac.caltech.edu/data/Planck/release_2/ancillary-data/}}.
We find that the odd multipole alignment axes are stable up to an exclusion of $10\%$ of the sky in the galactic plane. However, the even multipole common axes are found to be sensitive to galactic cuts. They progressively move towards or away from the galactic plane in the varying $l_{max}$ and $l_{min}$ cases respectively,
while remaining broadly clustered. Applying a galactic mask with $80\%$ or less sky fraction is found to destroy the collective orientation of these axes. This analysis is presented in Appendix~\ref{apdx:skycuts}.

 We then tested the effect of including more multipoles by extending the
multipole range to $l_{max}=71,81,91$ and $101$. Any significant
alignments seen in studying the multipole window $l=[2,61]$ vanish.
This is not unexpected, as it could be a simple consequence of diluting the signal.

Finally, we analysed clean CMB maps obtained using other cleaning procedures
and data sets. We find a similar behaviour for the even/odd multipole common
axes in WMAP provided nine year Internal Linear Combination\footnote{\url{https://lambda.gsfc.nasa.gov/product/map/current/}}
(ILC) map \citep{wmap9yrfinalres}, and the Local-generalized Morphological Component Analysis (LGMCA)
map that was produced using both the WMAP and PLANCK
full mission observations\footnote{\url{http://www.cosmostat.org/product/lgmca_cmb/}} \citep{lgmca}.

We also checked collective alignment axes in multipole blocks of $\Delta l = 6$ from the same
range $l=[2,61]$, with three even/odd multipoles in each block. The alignment axes thus inferred
for even/odd multipoles accordingly span the same region, from lowest multipole bin to the highest,
as seen in varying $l_{max}$ and $l_{min}$ cases. However the cumulative statistics are better
suited for our purpose i.e., to probe the widest possible correlations across (even/odd) multipoles.

\subsection{Dissecting cumulative statistics}

The cumulative statistics do not give much information on which regions of the data dominate the analysis.  The Alignment entropy is also just a single-number summary that cannot completely identify the source of this anomaly.
To glean more information about the observed alignments, we look inside
the cumulative statistics in this section, while also introducing an {\it independent} statistic for testing isotropy.

To make a more informative statistic from PEVs $|\tilde{e}_l \rangle$, we first observe that normalized eigenvectors are equivalent to rank-1 projection operators $\Pi_l = | \tilde{e}_l \rangle \langle \tilde{e}_l |$. We can then define a \emph{Hilbert-Schmidt} inner product (HSIP) \citep{ReedSimon} as
\begin{equation}
B_{ll'} = Tr\{\Pi_l^\dagger \Pi_{l'}\} = \langle \tilde{e}_l | \tilde{e}_{l'} \rangle^2.
\end{equation}

 For a set of `$n$' unit vectors, there will be a total of `$n(n-1)/2$' such independent inner products possible. The distribution of these independent HSIPs treated as a random variable, $B_{ll'} \rightarrow x$ (for all $l$, and $l'<l$), has an analytic form given by $f(x)=1/(2\sqrt{x})$  for $0 \leq x \leq 1$ (see Appendix \ref{apdx:null} for details). Correspondingly, its cumulative
distribution function is given by $F(x)=\sqrt{x}$.  We refer to the analytic isotropic null
distribution function as \emph{aPDF}, and the corresponding cumulative distribution
function as \emph{aCDF}. Analogously, we refer to the empirical counterparts
as \emph{ePDF} and \emph{eCDF}, respectively. The aPDF in this form is normalized
to have unit area under the curve.

\begin{figure}
\centering
\includegraphics[width=0.9\columnwidth]{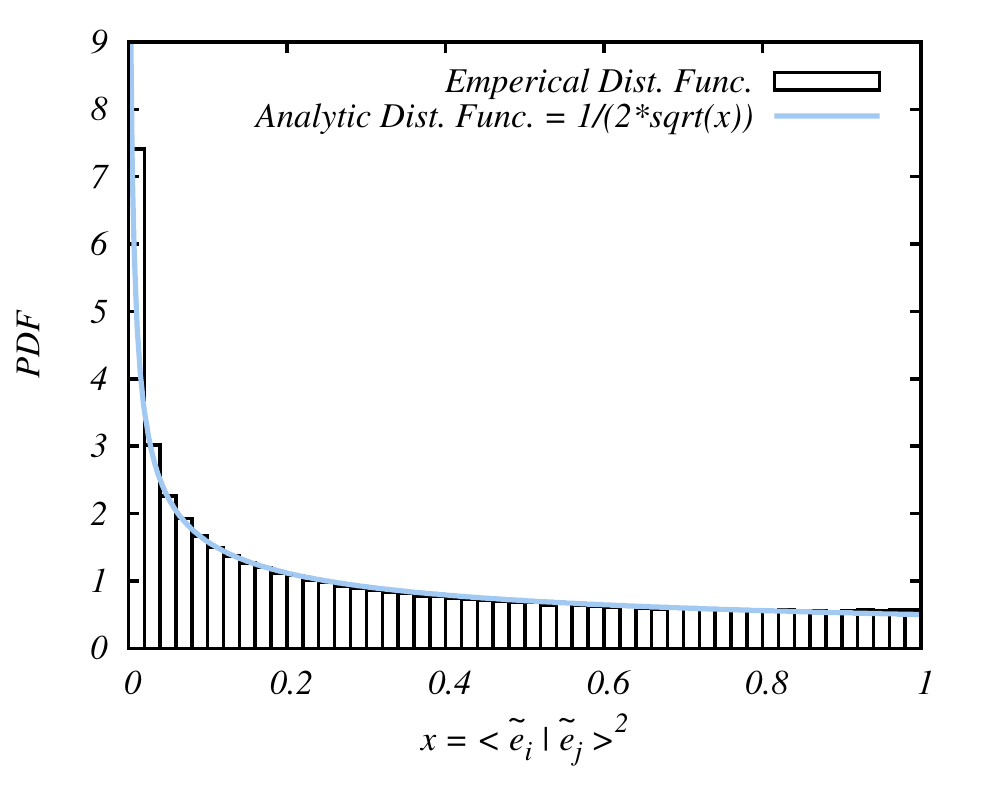}
\caption{Test of agreement between empirical PDF of Hilbert-Schmidt inner product of
         isotropically distributed unit vectors on a sphere, and their analytic
         distribution function. The simulation used 1000 random sets of
         $30$ isotropically distributed unit vectors recording the ePDF
         histogram each time. The mean ePDF  obtained from averaging individual ePDFs
         is shown here in bars. The analytic PDF is shown as a solid (blue) line.}
\label{fig:apdf-epdf}
\end{figure}

 Before proceeding further we first check that $f(x)=1/(2\sqrt{x})$ is the true PDF of Hilbert-Schmidt inner products of isotropically distributed unit vectors. We generate 1000 sets of $n=30$ units vectors. All possible HSIPs among
these unit vectors are computed for each set of $30$ normalized vectors which will be a total of $30 \times 29/2=435$. Then the \emph{mean} empirical distribution function is built by taking the average of individual ePDF histograms of 1000 sets of $30$ isotropic unit vectors to compare with the analytic distribution function. The \emph{mean} and \emph{analytic} PDFs are shown in Fig.~[\ref{fig:apdf-epdf}].
The $435$ independent HSIPs for each set of $30$ isotropic unit vectors are sorted
into $50$ bins to compare the aPDF and ePDF. We find  excellent agreement between
the two distribution functions.

\begin{figure*}
\centering
\includegraphics[width=0.9\textwidth]{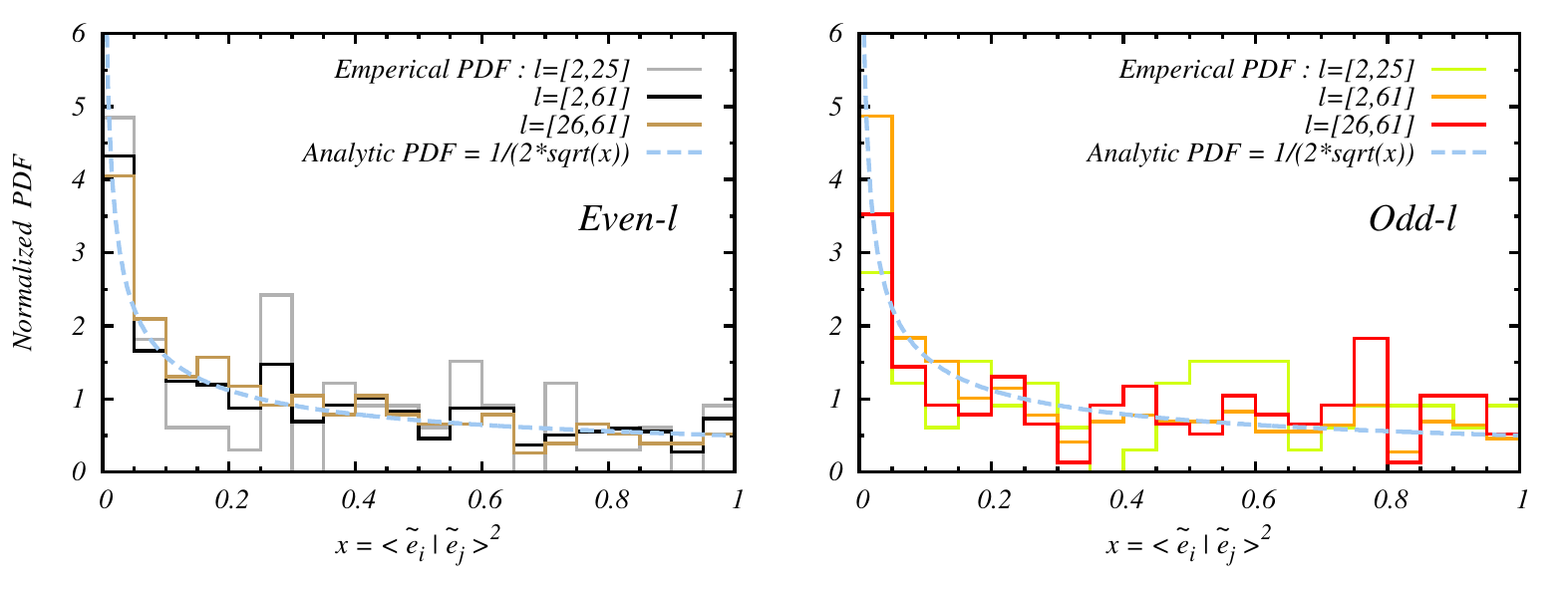}
\caption{The empirical distribution functions of Hilbert-Schmidt
         inner products (HSIPs) of PEVs from data computed separately for even and odd
         multipoles, from three representative multipole ranges $l=[2,25]$, $[2,61]$
         and $[26,61]$ are shown here. The even and odd HSIP ePDFs are shown in \emph{left}
         and \emph{right} panels respectively. The analytic distribution function is shown
         by a dashed (blue) line.}
\label{fig:data-epdf}
\end{figure*}

\begin{figure*}
\centering
\includegraphics[width=0.9\textwidth]{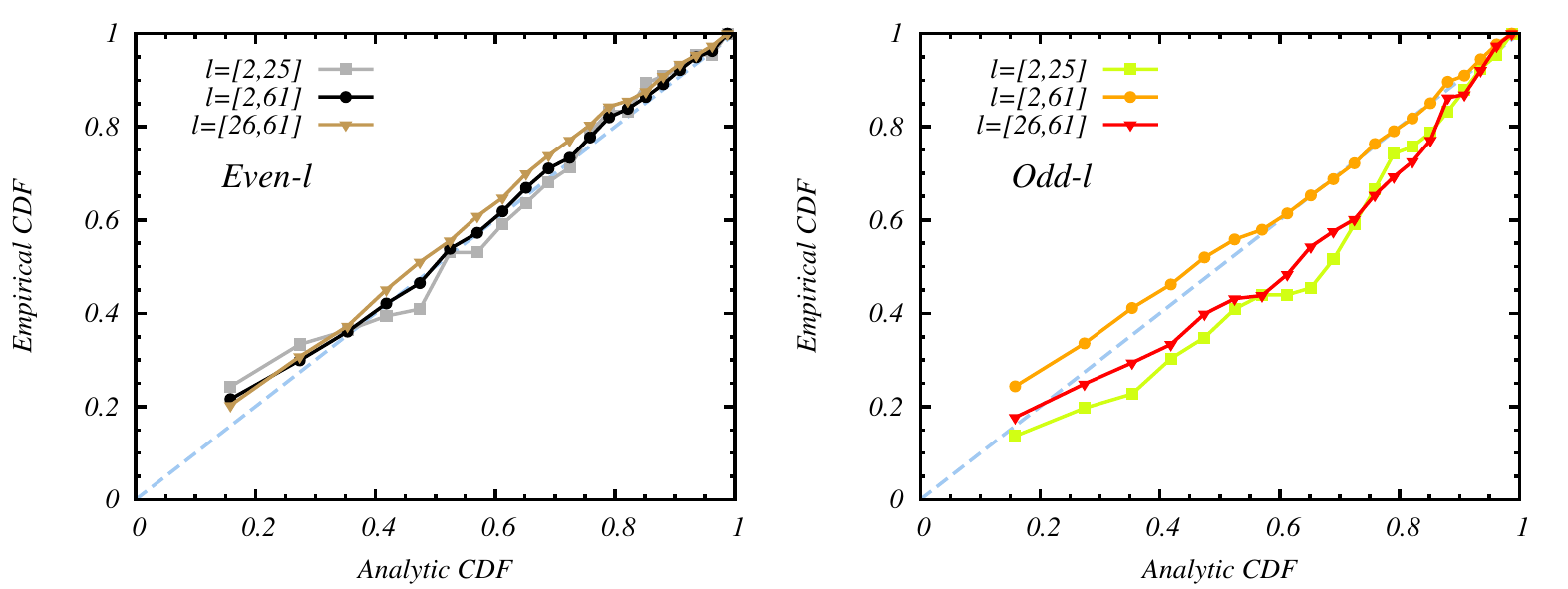}
\caption{Same as Fig.~[\ref{fig:data-epdf}], but shown here are the empirical \emph{cumulative}
         distribution functions built from data HSIPs. See text for details.}
\label{fig:data-ecdf}
\end{figure*}

Now we evaluate the ePDF and eCDF of HSIPs from the data (PLANCK 2015 \texttt{Commander} map)
and compare them with their analytic forms.
We illustrate the distributions for three representative multipole ranges
$l=[2,25]$, $[2,61]$ and $[26,61]$. There are a total of $12$, $30$ and $18$
even or odd multipole PEVs in these three sets. Thus $12 \times 11/2=66$,
$30 \times 29/2 = 435$ and $18 \times 17/2 = 153$ independent HSIPs
are possible, respectively, in each set of multipoles among even or odd
multipole PEVs. Recall that, in the cumulative statistics, we chose the multipole
range such that there are equal number of even/odd multipoles available in the $l-$range
being considered. These are then sorted into $20$ bins to build the ePDF and eCDF.
The results are shown in Fig.~[\ref{fig:data-epdf}] and [\ref{fig:data-ecdf}], for the
three multipole ranges mentioned above. In describing these plots below, we only highlight
a \emph{visual} discrepancy. Later, we use \emph{Anderson-Darling} (AD) test statistic
to find whether the data conforms with the isotropic null distribution function or not,
and also quantify it's significance using simulations. 

The eCDF plots highlight the peculiarity of odd multipole PEV alignments rather more dramatically than ePDF plots. One notices that there is a mild deficit at low HSIP
bin values, and a mild excess at intermediate HSIP bin values in the empirical
PDF of odd multipole PEV alignments for the range $l=[2,25]$
in Fig.~[\ref{fig:data-epdf}]. The discrepancy with the isotropic hypothesis is
more striking in the empirical cumulative distribution function of odd multipole
PEV HSIPs for the same range
compared to the analytic distribution in Fig.~[\ref{fig:data-ecdf}].
With larger $l_{max}=61$, the discrepancy nearly vanishes.
The diagonal dashed line is the reference curve about which the data
statistic coming from the null distribution is expected to fluctuate. The empirical CDF of even multipole PEV HSIPs
essentially criss-crosses this reference curve in Fig.~[\ref{fig:data-ecdf}],
in agreement with our findings from previous sections. However, as noted above,
the odd multipole alignments deviate significantly. Our earlier observation on the
presence of two populations of anisotropy axes is also corroborated by the eCDF
curves for $l=[2,25]$ and $l=[26,61]$ that are non-overlapping in multipole range.

The \emph{Anderson-Darling} ($AD$) test \citep{ADstat,BohmZech} quantifies the agreement of the data with
the isotropic null distribution. The Anderson-Darling statistic is defined as
\begin{eqnarray}
AD &=& - N - \sum_{i=1}^N \frac{2i-1}{N} \left[ \ln(F(x_i)) + \right. \nonumber \quad\quad \\
     && \quad\quad \left. \ln(1-F(x_{N-i+1})) \right]\,,
\end{eqnarray}
where `$N$' is the number of sample points, and $F(x_i)$ is the analytic cumulative distribution
function evaluated for the data sample point $x_i$. For our specific case of HSIPs,
$F(x_i) = \sqrt{x_i}$, and for a set of `$n$' even/odd multipole PEVs, there are
$N=n(n-1)/2$ number of independent inner products possible.
Similar to the case of varying $l_{max}$ discussed in the previous section,
the $AD$ statistic is obtained as a function of $l_{max}$ from the
multipole range $l=[2,61]$. At each  $l_{max}$, the $AD$ statistic is
computed from the even/odd multipole PEV sub sets of the current $l-$range separately.
Likewise, we also show the results for varying $l_{min}$ case.

\begin{figure*}
\centering
\includegraphics[width=0.9\textwidth]{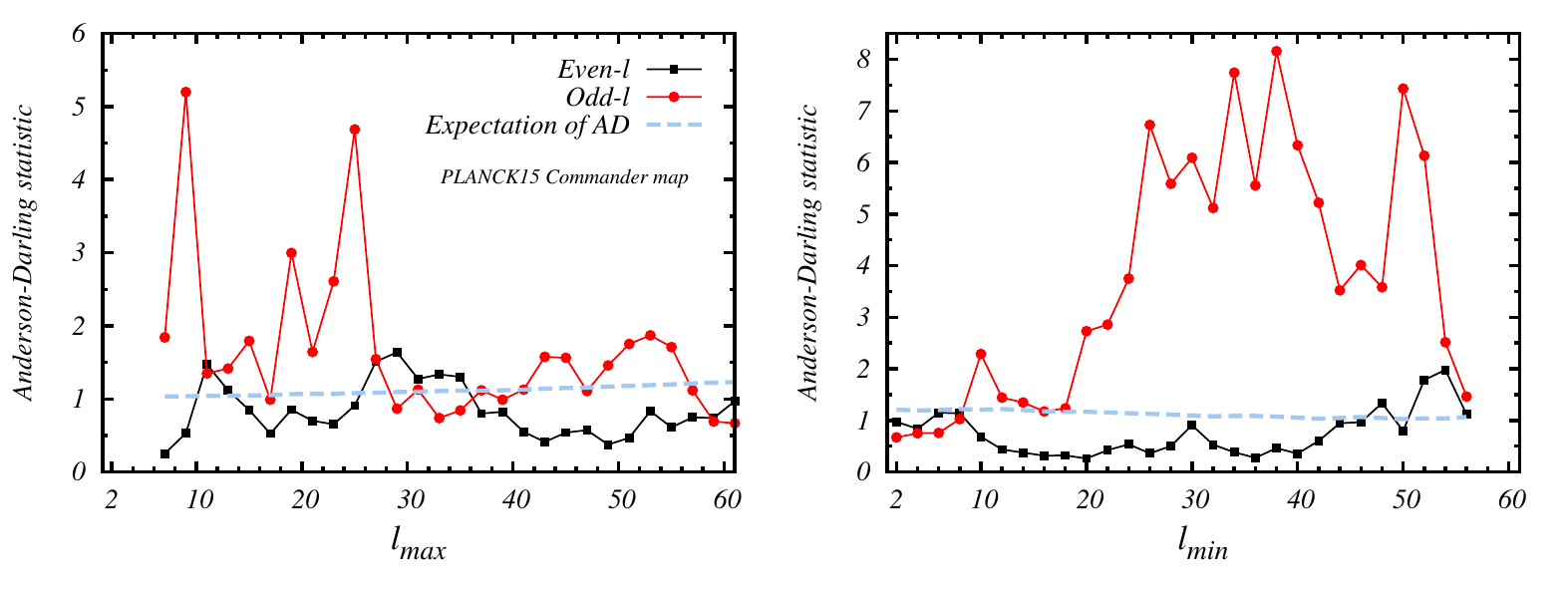}
\caption{The Anderson-Darling statistic computed separately for even and odd multipole
         PEV HSIPs from PLANCK 2015 \texttt{Commnader} map are shown here.
         The case of varying $l_{max}$($l_{min}$) are shown in \emph{left}(\emph{right})
         panel. The even and odd multipole statistic values are
         shown in black and red solid lines with square and filled circle point types
         respectively. The dashed (blue) curve denotes the expected statistic value, obtained
         from an ensemble of 1000 mock observed CMB maps, that is same for even
         or odd multipole PEVs.}
\label{fig:data-AD-stat}
\end{figure*}

 The $AD$ statistic values as a function of $l_{max}$ are shown in the left panel of Fig.~[\ref{fig:data-AD-stat}], and as a function of $l_{min}$ in the right panel
in the same figure. The expected value of the $AD$ statistic 
is denoted by a (blue) dashed line. It is computed from 1000 ILC-like noisy
CMB maps obtained from FFP simulations described in Sec.~[\ref{sec:data-sim}].
The mean $AD$ statistic from simulations is evaluated in  both cases
for even and odd multipoles separately. Since the two curves are indistinguishable,
as expected, only one of them is shown to avoid redundancy. From Fig.~[\ref{fig:data-AD-stat}],
one can readily see that the $AD$ statistic for $l_{max}=9,19,23$ and $25$ acquires
very high values, hinting at the origin of the $~2\sigma$ level significance
seen for the common alignment axes of odd multipole PEVs on large angular scales.
From Fig.~[\ref{fig:plk15cmdr-varlmin}], we see that many of the collective alignment
axes in the case of varying $l_{min}$ settled in the galactic plane. Correspondingly,
in the right-hand panel of Fig.~[\ref{fig:data-AD-stat}], we see that the distribution of the HSIPs quantified by the $AD$ statistic is very high compared to its expectation in the same multipole range,
in the varying $l_{min}$ case.

\begin{figure*}
\centering
\includegraphics[width=0.9\textwidth]{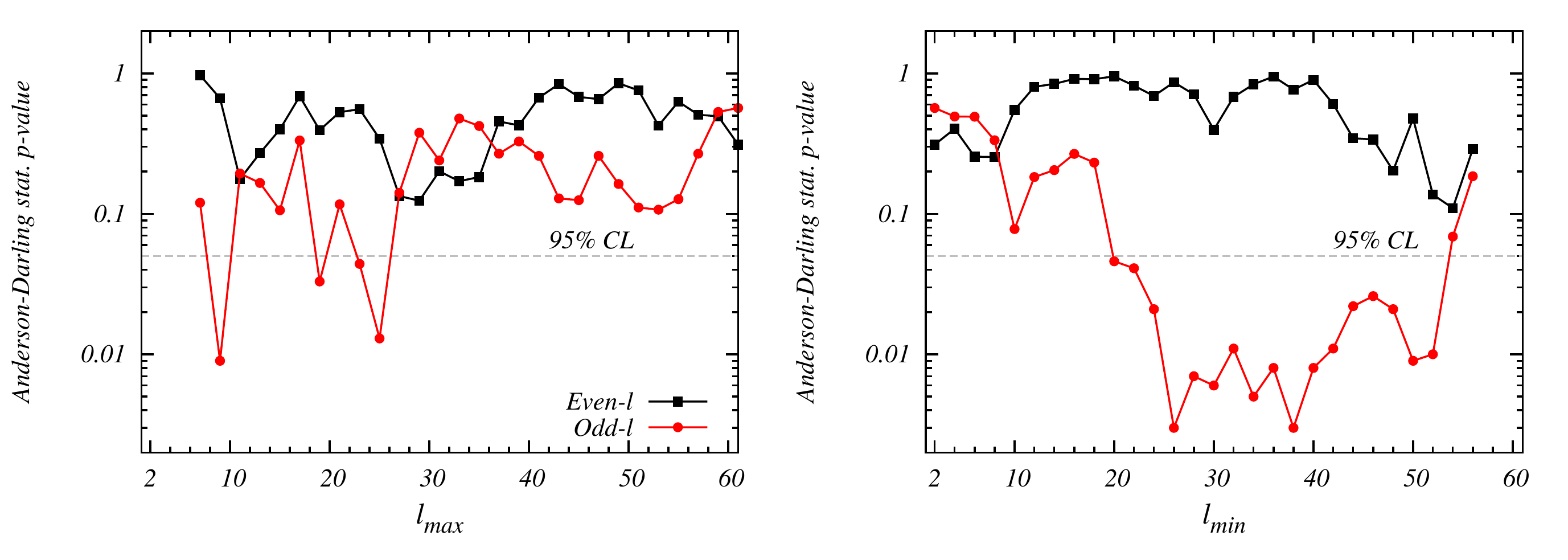}
\caption{$p-$values of AD statistics of data shown in Fig.~[\ref{fig:data-AD-stat}],
         are plotted here as a function of $l_{max}$ and $l_{min}$. The significances show a
         similar trend for even or odd multipoles as seen with the Alignment tensor
         method in Fig.~[\ref{fig:plk15cmdr-varlmax-signf}] and [\ref{fig:plk15cmdr-varlmin-signf}].
         The $95\%$ confidence level is also shown for reference, as a dashed grey line in the plot.
         The statistic shows higher significances for $l_{max}=9,19,23,25$ indicating
         the possible source of the $\sim2\sigma$ significances seen earlier,
         with odd multipole PEV alignments on large angular scales, in the Alignment
         entropy analysis.}
\label{fig:pval-AD-stat}
\end{figure*}

 The $p-$values of the $AD$ statistic
for the PLANCK 2015 \texttt{Commander} map derived HSIPs as a function of $l_{max}$
are shown in the left-hand panel of Fig.~[\ref{fig:pval-AD-stat}].
The significances of the $AD$ statistic for even and odd multipole PEV HSIPs are computed separately, and are shown in black and red solid lines with  square and circle point types respectively.

 The Anderson-Darling statistic gives independent confirmation that the odd multipole PEV alignments are anomalous on large angular scales. Significance exceeding $2\sigma$ confidence level is found for 
$l_{max}=9,19,23$ and $25$ which are found to have high values for $AD$ statistic
from the left plot of Fig.~[\ref{fig:data-AD-stat}].
The even multipole PEV HSIPs show no significant signal of differing from the isotropic null distribution in this analysis,
consistent with the finding from
preceding section. Thus there
are some anomalous alignments among odd multipole anisotropy axes on large
angular scales represented by their principal eigenvectors that are
resulting in the high significance of our test statistic.
Owing to the highly deviant $AD$ statistic in the varying $l_{min}$ case, the $AD$ statistic is found to be anomalous for the same range of multipoles. The $p-$value plot for the same is shown in right panel of Fig.~[\ref{fig:pval-AD-stat}], which follows a
trend similar to the significances found in Fig.~[\ref{fig:plk15cmdr-varlmin-signf}].

\section{Conclusions}

We have compared alignment statistics of parity even and odd multipoles with several independent methods. We used the clean CMB signal estimate from PLANCK 2015 data obtained using the
\texttt{Commander} algorithm. Analysis was restricted to the first sixty multipoles i.e., $l=[2,61]$. Power tensor and Alignment tensor
statistics were used to probe the alignments of even
and odd parity multipoles, separately.

We studied the data in several ways. The collective alignment axes of even and odd multipoles show different behaviors. The anisotropy axes of even-parity  multipoles
from large angular scales  are broadly clustered near the direction of the CMB dipole.  The anisotropy axes of odd multipoles are much less concentrated, but are significantly directional as quantified by Alignment entropy.

We constructed cumulative statistical measures that fixed the lower limit $l_{min}=2$, while varying the upper limit to reach $l_{max}=61$.  The Alignment entropy, $S_X$,	 of even-parity multipoles was as expected from an uncorrelated isotropic distribution. The odd--parity multipole $S_X$ was unusually small
on large angular scales with significance exceeding $2\sigma$ magnitude.  As $l_{max}$ was increased above $l_{max}\sim 27$  the significance disappeared, apparently by dilution in the larger set.  This significance
nevertheless disappears by ignoring the first few multipoles. A similar effect was seen in studying even-odd multipole power asymmetry, using the WMAP seven year temperature power spectrum \citep{Aluri12}.
To understand the alignment preferences of small angular scales in the range being studied,
we fixed the upper limit at $l_{max}=61$ while varying the lower limit $l_{min}$.
A regime of multipoles with small $S_X$ at $2\sigma$ or more significance for odd-parity multipoles was observed, with lowest $p-$value for $S_X$ occurring at $l_{min} \gtrsim 26$. The two different effects from varying $l_{min}$ and $l_{max}$ analysis in a single data set pose a puzzle.  The resolution may involves two different populations separated by a middle range of $l\sim 27$, with each population diluting a distinctive signal of the other when populations are mixed. 
The observation that the axes of the $ l>27$ set settled at the galactic plane may be an indication of a residual galactic bias in this subset.

These results are further tested against potential residual contamination in the full sky map
by  excising different fractions of the sky, and then inpainting the masked region.
 The odd multipoles' common axes are stable against
galactic cuts up to excluding (and then inpainting) $10\%$ of the sky, whereas the even multipoles are found to be sensitive to galactic cuts.

An independent statistic was used to dissect the cumulative statistical studies. 
The Hilbert-Schmidt inner products (HSIP) are rotationally invariant statistics with an analytic isotropic null distribution. The distribution of the data compared to the HSIP null was computed using Anderson-Darling (AD) test statistic. 
For the odd multipole PEVs, the AD statistic
for the data HSIPs shows a significance similar to that found using
the Alignment entropy method. The AD method pinpoints $l_{max}=9,19,23$
and $25$ as containing unusual alignments that are rendering the AD statistic
anomalous at a significance of $2\sigma$ or more. 

Interestingly, we find that the even mirror parity axis from the
PLANCK 2015 results, and the even multipoles' common axes from large angular scales
computed here, broadly point in the CMB dipole direction. Likewise,
the odd mirror parity axis from the PLANCK 2015 analysis, and the odd parity
low$-l$ hemispherical power asymmetry axis fall in the region spanned by the odd
multipole alignment axes. From these observations, we speculate that these anomalous
axes may have a common origin in their peculiar parity (a)symmetry properties.

We plan to investigate these speculations more in a later work.

\section*{Acknowledgements}

We acknowledge the use of freely available
\texttt{HEALPix}\footnote{\url{http://healpix.jpl.nasa.gov/}}~\citep{Healpix} package and
\texttt{iSAP} software\footnote{\url{http://www.cosmostat.org/software/isap/}}
in this work.
Part of the results presented here are based on observations obtained with
PLANCK\footnote{\url{http://www.esa.int/Planck}}, an ESA science mission with instruments
and contributions directly funded by ESA Member States, NASA, and Canada.
We also acknowledge the use of WMAP data made available from Legacy Archive for Microwave
Background Data Analysis\footnote{\url{https://lambda.gsfc.nasa.gov/product/map/dr5/}} (LAMBDA) site
that is a part of NASA's High Energy Astrophysics Science Archive Research Center (HEASARC).
This research used resources of the National Energy Research Scientific Computing (NERSC)
Center, a DOE Office of Science User Facility supported by the Office of Science of the
U.S. Department of Energy under Contract No. DE-AC02-05CH11231.\\
PKA is funded by the post-doctoral fellowship program of the Claude Leon
Foundation, South Africa at UCT.
This work is based on the research supported by the South African Research Chairs
Initiative of the Department of Science and Technology and the National Research Foundation
of South Africa as well as the Competitive Programme for Rated Researchers (Grant Number 91552) (AW).
Any opinion, finding and conclusion or recommendation expressed in this material is that
of the authors and the National Research Foundation (NRF) of South Africa does not accept
any liability in this regard.\\
PKA also thanks Pankaj Jain for helpful exchanges on an earlier version of the paper.
AW would like to thank David Spergel for helpful discussions on this work.\\
We thank the anonymous referee for a careful reading and helpful comments on our paper.




\appendix

\section{Stability of alignment axes}
\label{apdx:skycuts}

\begin{figure}
\centering
\includegraphics[width=0.9\columnwidth]{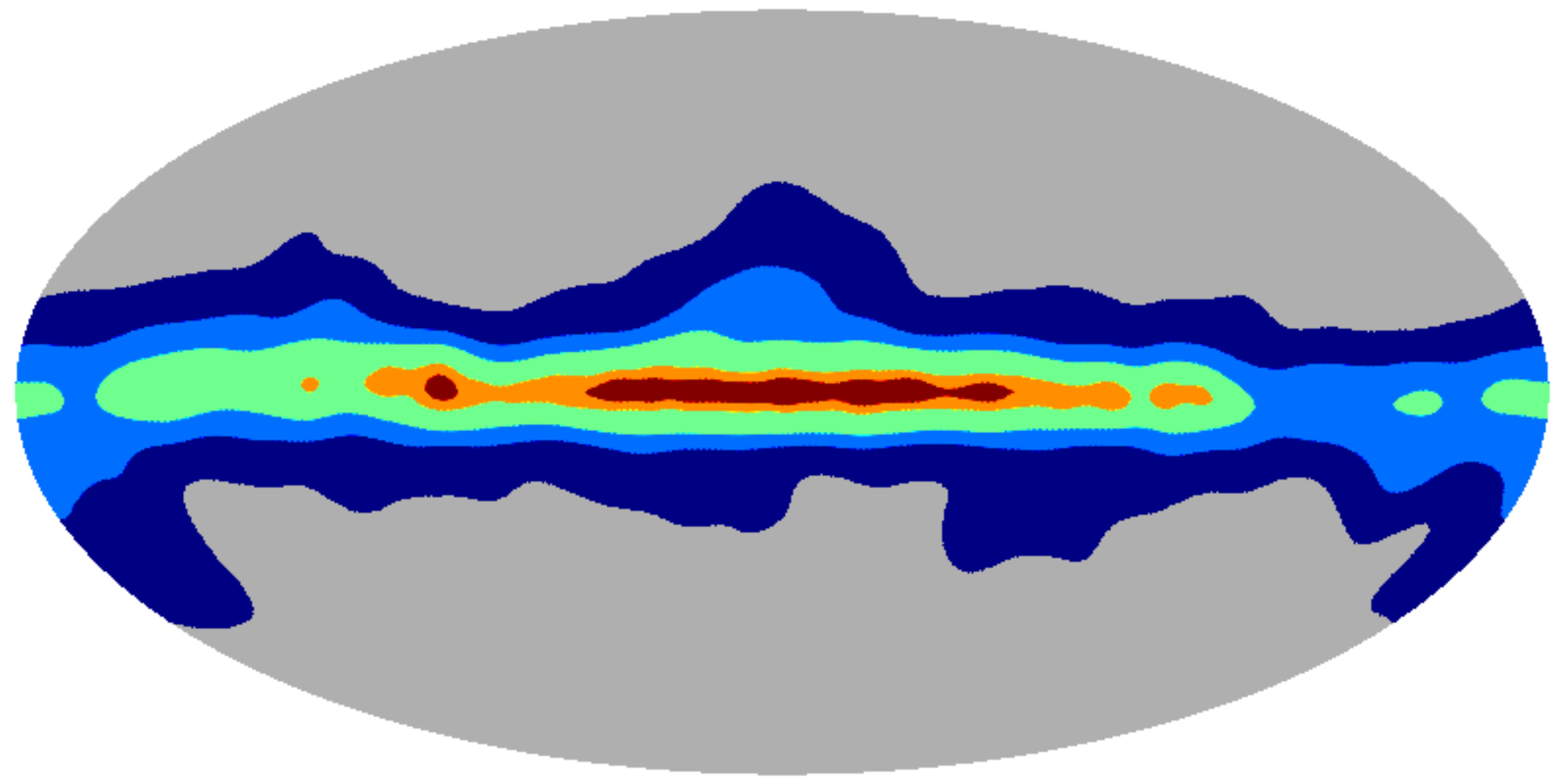}
\caption{Foreground exclusion masks that are applied to test the stability of the collective
alignment axes. From red to deep blue they progressive exclude 1\%, 3\%, 10\%, 20\% and 30\% of the sky.
We used the freely available iSAP software to inpaint the masked region.}
\label{fig:gal-masks}
\end{figure}

Here we probe the stability of the even/odd multipole alignment axes using different
foreground exclusion masks.
We used PLANCK 2015 HFI masks with varying sky fractions, that are provided
along with the second public release of PLANCK data\footnote{\url{http://irsa.ipac.caltech.edu/data/Planck/release_2/ancillary-data/}}.
The respective sky fractions of the masks used are 1\%, 3\%, 10\%, 20\% and
30\%. The excluded regions corresponding to these masks are shown in Fig.~[\ref{fig:gal-masks}].

We used these masks at their native resolution of \texttt{HEALPix} $N_{side}=2048$
on the PLANCK 2015 \texttt{Commander} CMB temperature map which is also made available at the
same resolution. The masked CMB map is then inpainted using the freely available
\texttt{iSAP} software\footnote{\url{http://www.cosmostat.org/software/isap/}}
(see \cite{inpaint}). We used the default settings of the
{\verb+mrs_alm_inpainting+} facility of \texttt{iSAP} to inpaint the CMB sky.

Following the same procedure as described in the main analysis,
the inpainted CMB map is then downgraded to $N_{side}=256$ and simultaneously
smoothed to have a beam beam resolution of $FWHM=1^\circ$ (degree) Gaussian beam.

\begin{figure*}
\includegraphics[width=0.48\textwidth]{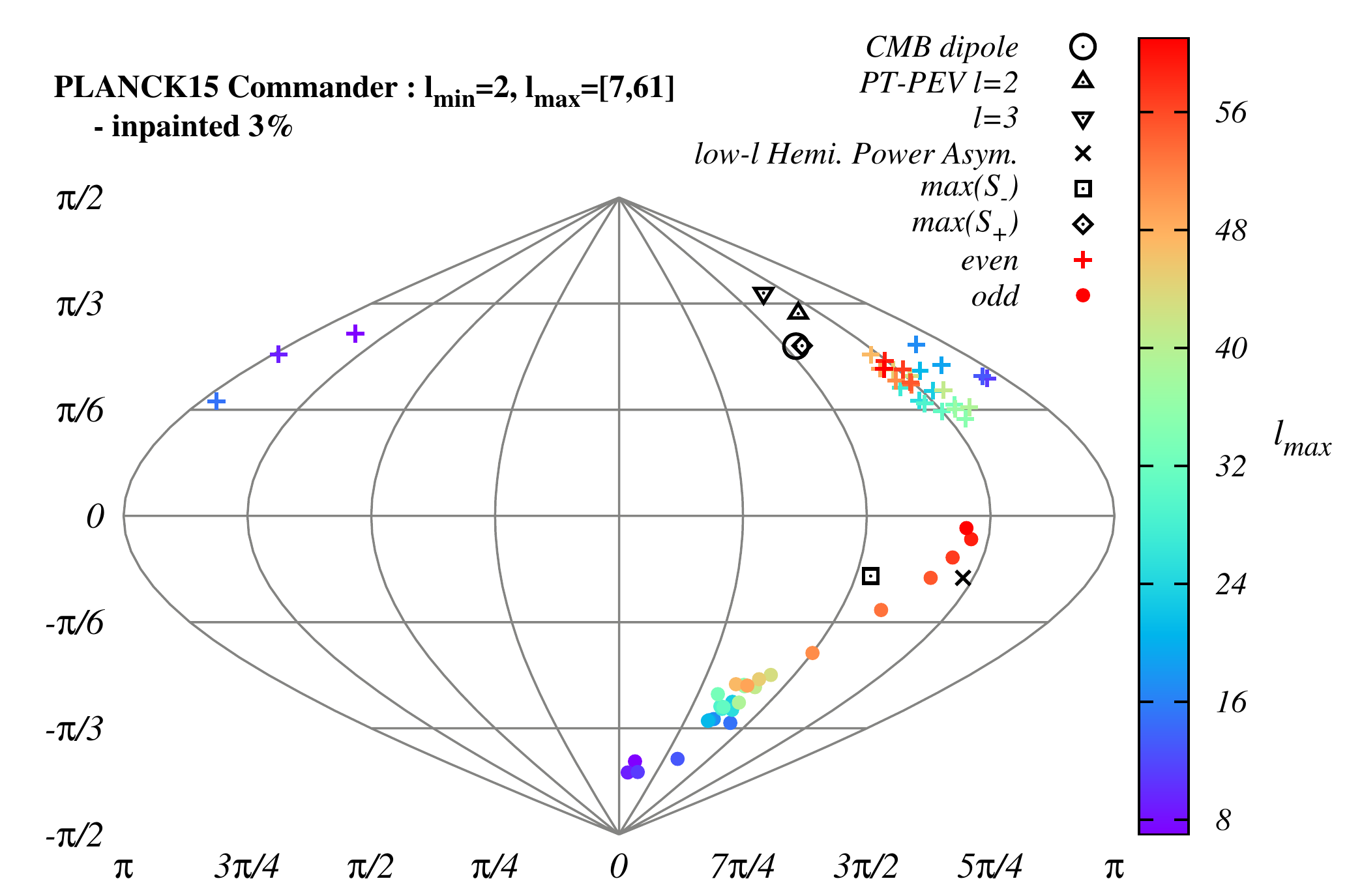}
~
\includegraphics[width=0.48\textwidth]{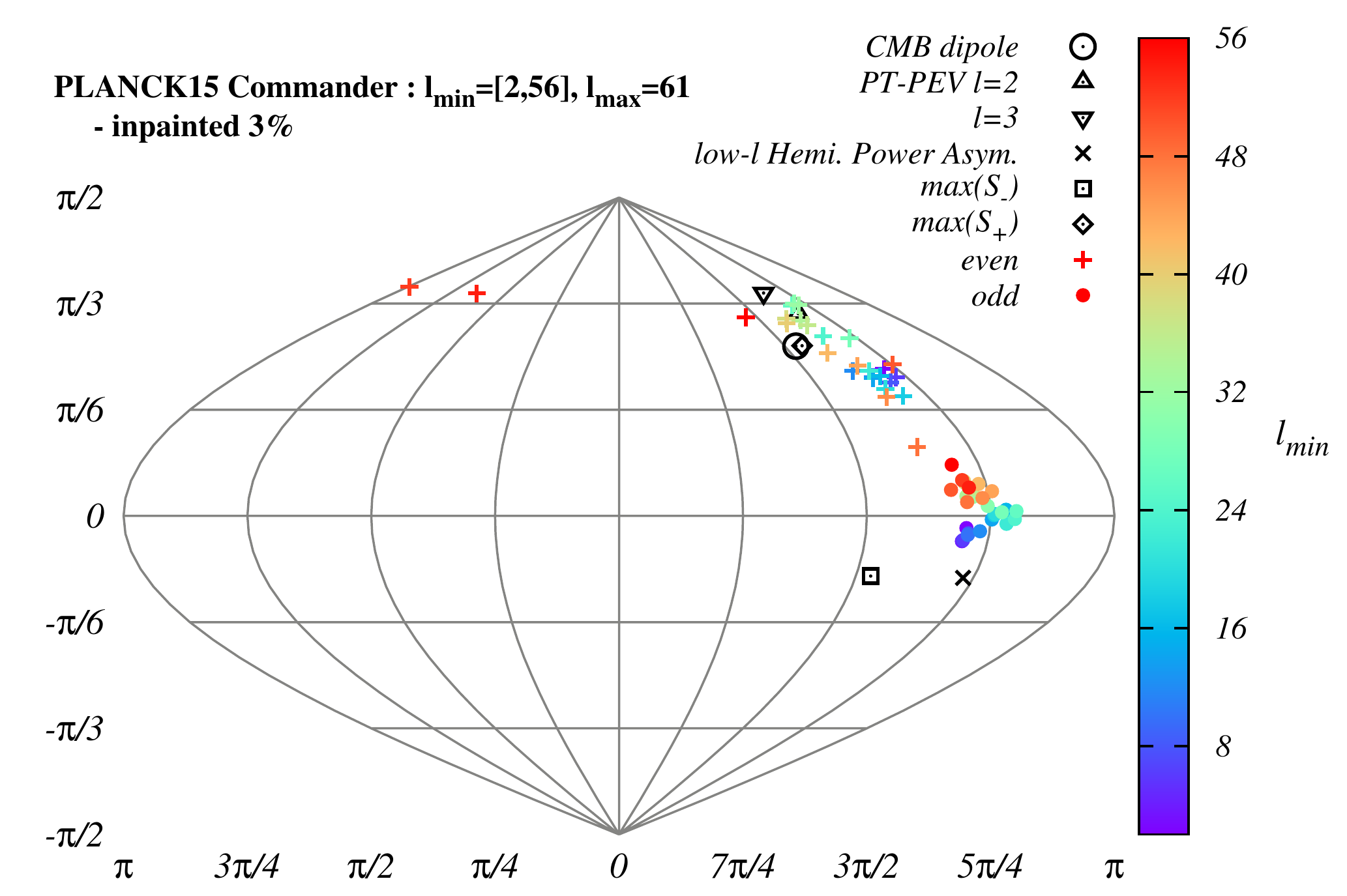}

\vspace{1.5em}

\includegraphics[width=0.48\textwidth]{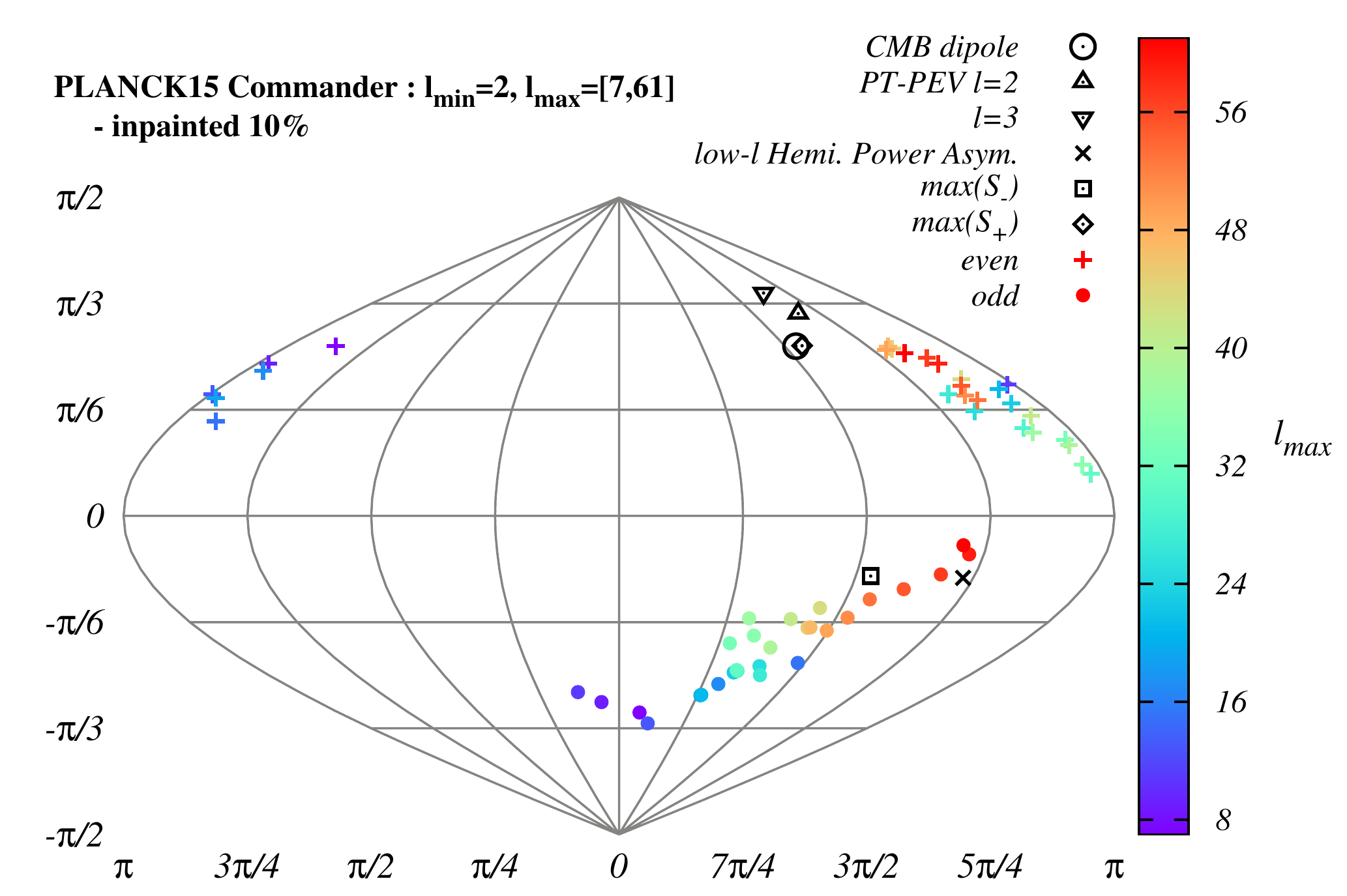}
~
\includegraphics[width=0.48\textwidth]{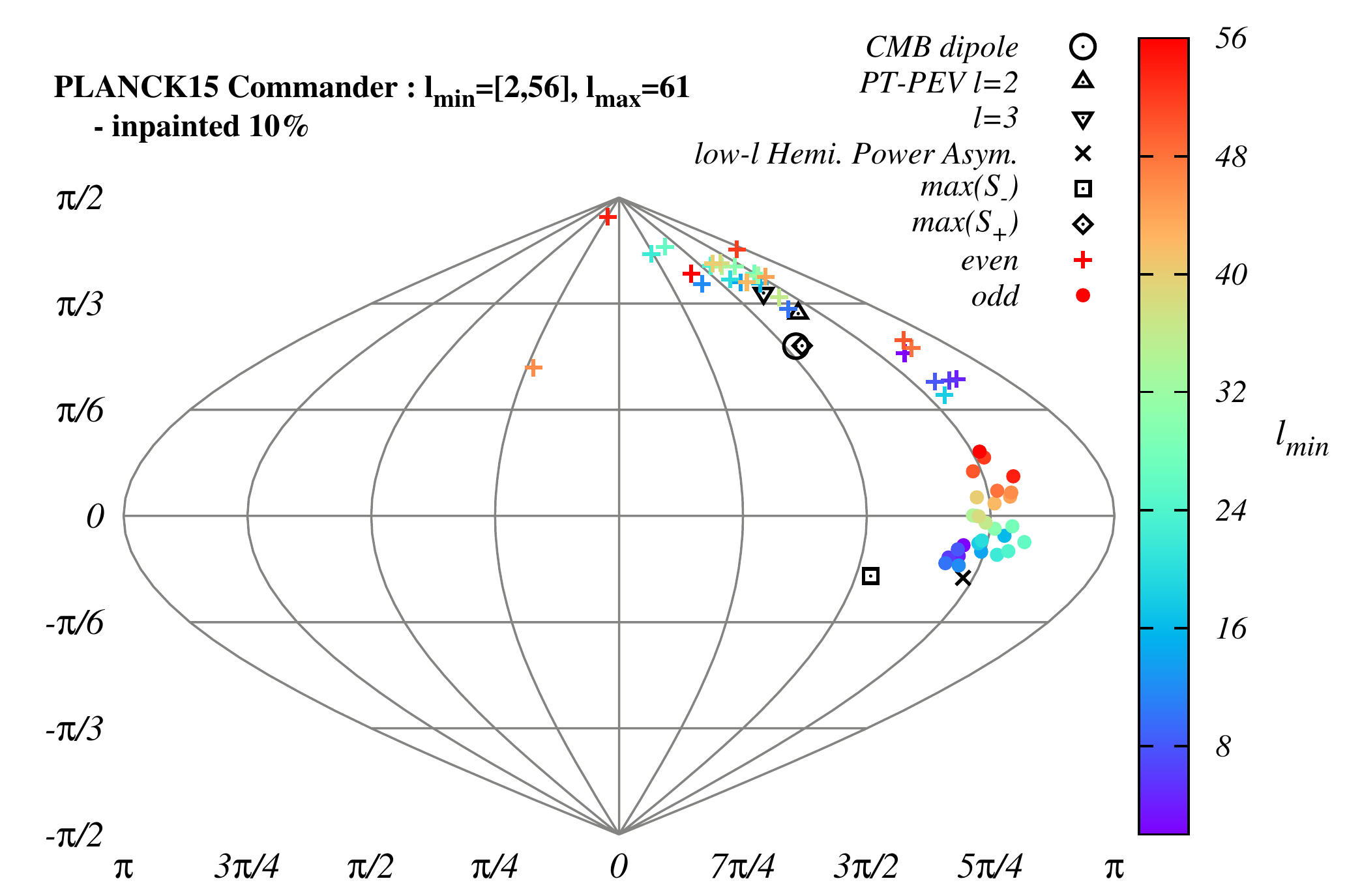}

\vspace{1.5em}

\includegraphics[width=0.48\textwidth]{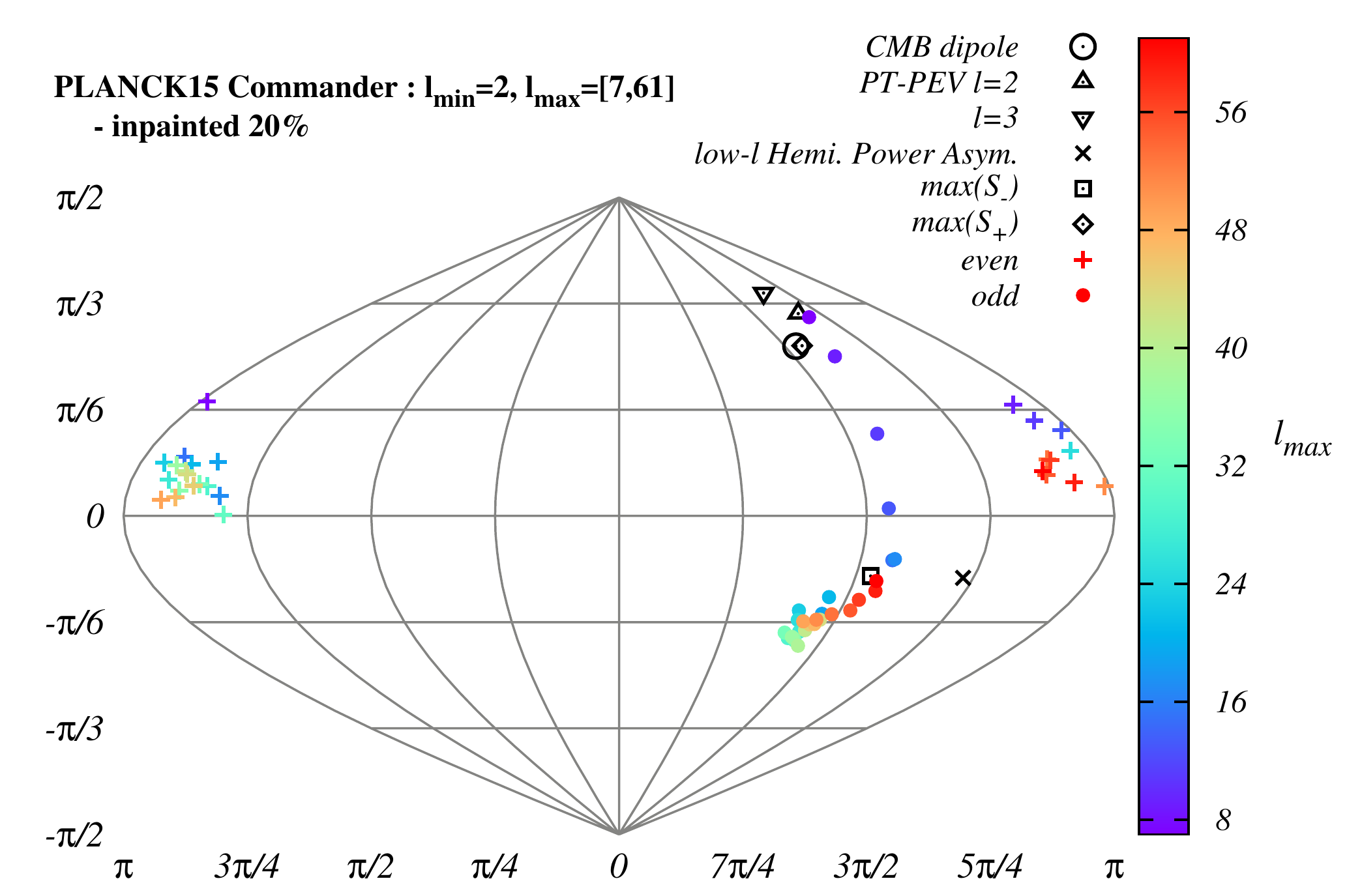}
~
\includegraphics[width=0.48\textwidth]{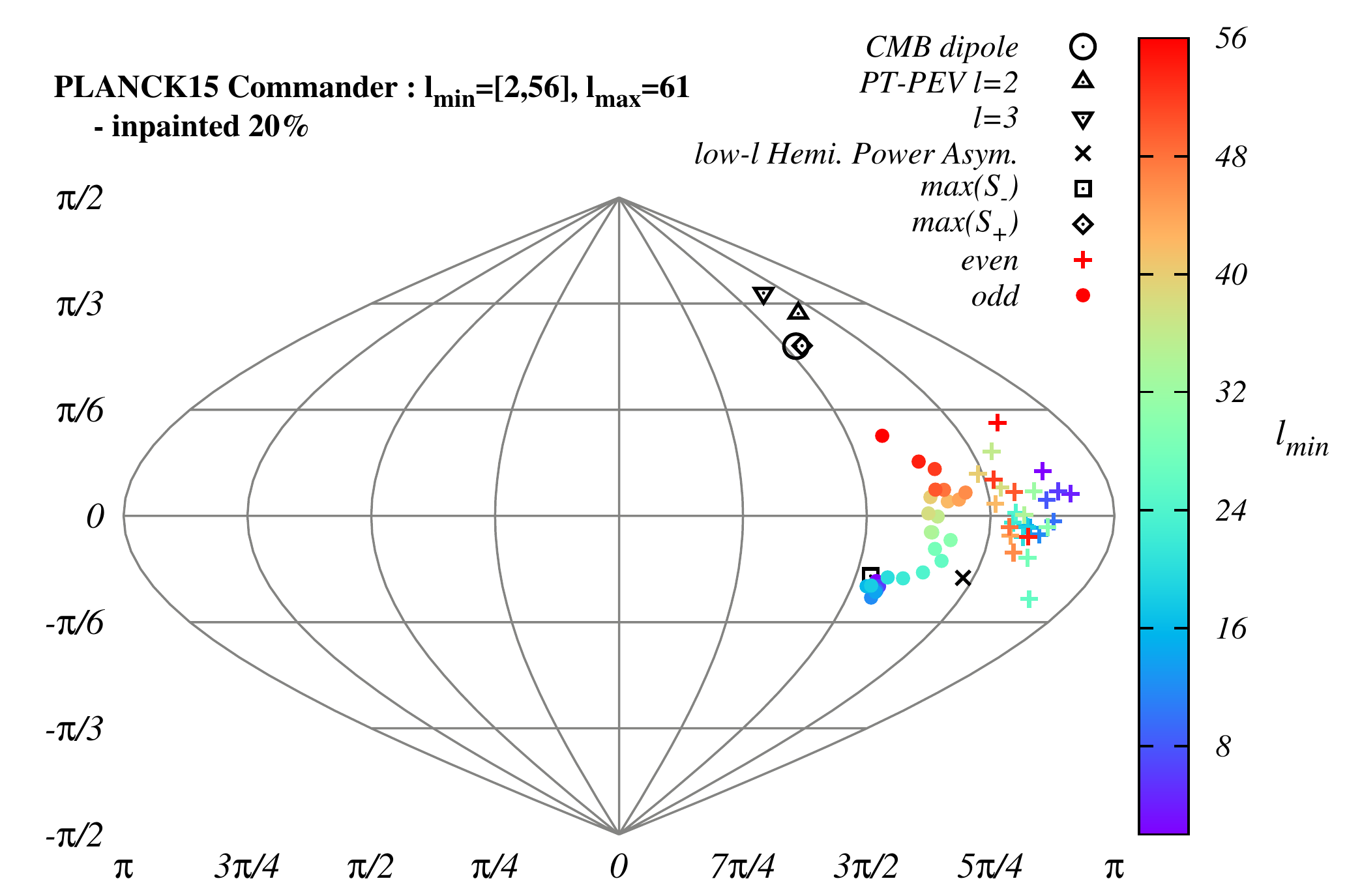}
\caption{Common alignment axes obtained after applying galactic masks with different sky
fraction and inpainting using \texttt{iSAP}. The varying $l_{max}$ and $l_{min}$ cases are shown in
the \emph{left} and \emph{right} columns, respectively, for masking fractions of 3\%, 10\% and
20\% of the sky.
By excluding 20\% or more sky fraction (and then inpainting), the broad orientations of the
common alignment axes disappears.}
\label{fig:inp-anls}
\end{figure*}

The common alignment axes of even and odd multipole PEVs from masking and inpainting
3\%, 10\% and 20\% of the CMB sky are shown in Fig.~[\ref{fig:inp-anls}].
Here we performed a qualitative analysis only. By visual inspection we see that
the odd multipole alignment axes are broadly stable up to 10\% of the sky being
masked and inpainted. However the even multipole PEV alignment axes steadily drift
towards galactic plane in the varying $l_{max}$, and move towards the poles in the
case of varying $l_{min}$. Applying galactic cuts with 20\% or more masking fraction
(followed by inpainting the masked sky) is found to destroy the alignment patterns seen otherwise.

\section{The Isotropic Null $HSIP$ Distribution}

\label{apdx:null}

Let $| \tilde{e}_{l} \rangle$ be a random eigenvector from an isotropic distribution. Since eigenvectors have no magnitude and no sign, $| \tilde{e}_{l} \rangle$ is equivalent to the rank-one projector $\Pi_l =| \tilde{e}_{l} \rangle \langle \tilde{e}_{l} |$. Consider the distribution of $x = Tr\{\Pi_l^\dagger \Pi_{l'}\} = \langle \tilde{e}_l | \tilde{e}_{l'} \rangle^2$ (for all $l \neq l'$). Choose coordinates where the first instance $| \tilde{e}_{1} \rangle$ is along the $z$ axis, so that $\langle \tilde{e}_l | \tilde{e}_{1} \rangle=\cos \theta_{l}$. In an isotropic ensemble the distribution of $\cos \theta_{l}$ is constant over the range $-1 \leq \cos\theta \leq 1$ as shown by the solid angle measure $d\Omega_{l}= d\cos \theta_{l} d \phi_{l}$. Averaging over all cases we can drop the index $l$. For each $x=\cos^{2}\theta$ there are two signs of $\cos\theta$. The distribution of $x$ over the range $0\leq x \leq 1$ is then \ba f(x)=\frac{dN}{dx} =2 \frac{dN}{d \cos \theta}| \frac{d\cos \theta}{dx}|= \frac{2}{2}\frac{d \sqrt{x}}{dx} = \frac{1}{2\sqrt{x}}. \nn \ea The same result comes from $ f(x)= (2\pi/ 4 \pi)\int_{-1}^{1} d \cos \theta \, \delta(\cos^{2}\theta -x)$, accounting for two solutions of the delta function.


\bsp	

\label{lastpage}

\end{document}